\documentclass{astlb}
\usepackage{graphicx} 
\usepackage{amsmath} 
\usepackage{epsfig,graphics} 
\usepackage{rotating} 
\usepackage{epstopdf}
\usepackage{pdfpages}
\usepackage{longtable}
\usepackage{caption}
\usepackage{dcolumn}
\usepackage{pdflscape}
\usepackage{hyperref}
\usepackage{comment}
\usepackage{color}
\usepackage{tcolorbox}
\usepackage{xspace}

\begin{document} 

\journalinfo{2021}{47}{9}{613}[633]
\title{X-ray observations of historical classical nova counterparts with \lowercase{e}ROSITA telescope aboard  SRG orbital observatory during the all-sky survey}

\author{\bf \hspace{-1.3cm} \copyright \ 2021 \
I. I.~Galiullin\address{1,3}\email{IlhIGaliullin@kpfu.ru}, 
M. R.~Gilfanov\address{2,3}
\addresstext{1}{Kazan Federal University, Kazan, Russia}
\addresstext{2}{Space Research Institute of Russian Academy of Sciences, Moscow, Russia}
\addresstext{3}{Max Planck Institute for Astrophysics, Garching b. Munchen, Germany}}
 
\shortauthor{Galiullin \& Gilfanov }  
\shorttitle{X-RAY OBSERVATIONS OF CLASSICAL NOVA COUNTERPARTS BY SRG/eROSITA} 
\submitted{05.08.2021;\ Accepted 05.08.2021}

\begin{abstract}  
X-ray emission from  counterparts of historical classical novae (CNe) in our Galaxy is
studied. To this end, we use data from three SRG/eROSITA sky surveys in the hemisphere analyzed by the RU eROSITA consortium.  Out of 309 historical
CNe, X-ray emission has been detected from 52 sources with 0.3–2.3 keV luminosities in the  $\rm L_X\approx10^{30}\sim 10^{34}$ erg/s range. Among them, two sources have supersoft spectra and are associated with the post-nova supersoft X-ray emission.  Hardness of  X-ray spectra of some of  the remaining sources tentatively suggests that magnetized white dwarfs (WDs) may account for some fraction of CN counterparts detected in X-rays. This hypothesis will be further tested  in the course of the  following SRG/eROSITA sky surveys. 

The CN counterparts represent a bona fide sample of accreting WDs with unstable  hydrogen burning on their surface, while their X-ray luminosity in quiescence is a reasonable proxy for the accretion rate in the binary system. Using this fact, we have constructed the accretion rate distribution of WDs with unstable hydrogen burning and compared
it with the accretion rate distribution of known steady supersoft X-ray sources in our Galaxy and nearby external galaxies. There is a pronounced dichotomy between these two distributions with the CN counterparts and the steady supersoft sources occupying different domains along the mass accretion rate axis, in accordance with the predictions of the theory of  hydrogen burning on the WD surface.

\end{abstract}

%----------------------------------------------------------------------------------------------

\section{Introduction}
The accumulation of accreting material on the surface of a white dwarf (WD) in a close binary system leads to thermonuclear hydrogen burning on its surface. Depending on the accretion rate and the WD mass, thermonuclear hydrogen burning can be non-steady or steady, leading to the phenomena of classical or recurrent novae \citep[e.g.,][]{1978ARA&A..16..171G,1982ApJ...257..752F,1982ApJ...257..767F,1989clno.conf...39S} and supersoft X-ray sources \citep[e.g.,][]{1992A&A...262...97V,2007ApJ...663.1269N,2013ApJ...777..136W}. Accreting WDs are one of the two main classes of candidates for type Ia supernova progenitors \citep[e.g.,][]{1973ApJ...186.1007W,1982ApJ...253..798N}. In the classical picture, the accumulation of material occurs in binary systems with one degenerate component during steady hydrogen burning on the WD surface, which can lead to an increase in the WD mass to the critical Chandrasekhar value $\rm \sim 1.4\ M_\odot$, to its collapse and thermonuclear explosion that is observed as a type Ia supernova explosion. The theory also predicts that at certain accretion rates and WD masses,  the WD mass can grow to the critical value  in the non-steady hydrogen burning regime as well \citep[e.g.,][]{2013IAUS..281..166S,2015MNRAS.446.1924H}.

Steady hydrogen burning on the WD surface occurs when the mass accretion rate from the companion star is close to $\dot{M}_{\rm acc}\sim 10^{-7}\ M_\odot$/yr (the exact values depend on the WD mass) and is accompanied by the release of an energy exceeding the gravitational energy of the accreting material by more than an order of magnitude. Since the energy release occurs deep in an optically thick layer of material, it leads to the generation of soft X-ray emission with a nearly black body spectrum and a temperature $\sim 10-100$ eV. Such sources with supersoft X-ray spectra and luminosities  of $\rm 10^{36}\sim10^{38 }\ erg/s$ (the so-called supersoft X-ray sources) were first discovered more than 40 years ago \citep[e.g.,][]{1981ApJ...248..925L,1991Natur.349..579T,1991A&A...246L..17G,1992A&A...262...97V}. Soon after their discovery, the emission from supersoft X-ray sources was successfully explained as the result of steady hydrogen burning on the surface of an accretion WD \citep[e.g.,][]{1992A&A...262...97V}. Owing to the observations of the ROSAT, Chandra, XMM-Newton X-ray observatories,  more than a hundred supersoft X-ray sources have been discovered in our and nearby galaxies \citep[e.g.,][]{2002ApJ...574..382S,2003ApJ...592..884D,2004ApJ...609..710D,2005A&A...442..879P,2021A&A...646A..85G}.

At lower accretion rates $\dot{M}_{\rm acc}\la 10^{-8}\ M_\odot$/yr, thermonuclear hydrogen burning becomes non-steady, leading to the phenomenon of classical novae (CNe) \citep[for a review, see, e.g.,  ][]{2020arXiv201108751C}. A total energy  $\rm \sim 10^{45}$ erg is released during a CN outburst. A nova outburst leads to ejecta into the interstellar medium \citep[see, e.g.,][]{1994ApJ...437..802K,2006ApJS..167...59H}. An X-ray emitting shock can be formed when the expanding ejecta interact with the interstellar medium or the wind from the donor star \citep[see, e.g.,][]{1977ApJ...213..492B}. The X-ray shock emission has a thermal spectrum with a temperature $kT\ga1$ keV and a luminosity $L_X\approx10^{33}-10^{35}$ \citep[see, e.g.,][]{1998ApJ...499..395B,2001A&A...373..542O}.

Thermonuclear burning of the material left after the CN outburst on the WD surface leads to the generation of a decaying supersoft X-ray emission, and a post-nova supersoft X-ray phase is observed once the ejecta have cleared \citep[see, e.g.,][]{2007ApJ...663..505N,2011ApJS..197...31S,2010A&A...523A..89H,2011A&A...533A..52H,
2014A&A...563A...2H}. The source spectrum during the post-nova supersoft X-ray phase is characterized by the absence of emission above $\ga$1 keV and, as a rule, is described by a black body model with a temperature $\la 70$ eV decreasing with time \citep[see, e.g.,][]{1993Natur.361..331O,1996ApJ...456..788K,2011ApJ...733...70N}. The duration of the post-nova supersoft X-ray phase depends on the accretion rate before the CN outburst and the WD mass \citep[see, e.g.,][]{2010AN....331..169H,2016MNRAS.455..668S}. The theoretical calculations of the relation between the duration and maximum effective temperature of the post-nova supersoft X-ray phase \citep{2013ApJ...777..136W} are consistent with the monitoring of CNe in the galaxy M31 \citep[][]{2010A&A...523A..89H,2011A&A...533A..52H,2014A&A...563A...2H}.

Once the post-nova supersoft X-ray emission has ended, the source returns to the “quiescent phase” in which the main energy source is the accretion of material from the companion star. The X-ray emission from cataclysmic variables in the quiescent phase is generated in the accretion disk, the boundary layer, and the optically thin, comparatively hot corona above the accretion disk \citep[see, e.g.,][]{1977MNRAS.178..195P,1981AcA....31..267T,1985ApJ...292..535P,1985ApJ...292..550P,
1988AdSpR...8b.135S}. In this regime, the CN counterparts manifest themselves as dwarf novae whose outbursts are associated with thermal instability in the accretion disk \citep[see, e.g.,][]{1984A&A...132..143M,1994A&A...288..175M}. The luminosity of cataclysmic variables outside dwarf nova outbursts characterizes the true accretion rate in such systems.

The Spektr--Roentgen--Gamma (SRG) orbital observatory was successfully launched from the Baikonur Cosmodrome on July 13, 2019 \citep{2021arXiv210413267S}. There are two grazing-incidence X-ray telescopes on board SRG: the ART-XC telescope named after M. N. Pavlinsky \citep{2011SPIE.8147E..06P} with the operating range  4–30 keV and the eROSITA telescope operating in the range 0.2–8 keV  \citep{2021A&A...647A...1P}. By mid-2021, the SRG observatory has completed three all-sky surveys.

In this work we study X-ray emission from sources associated with historical CNe in our Galaxy. The CN X-ray counterparts are confirmed cases of accreting WDs with unstable hydrogen burning on their surface. Below we will call these objects "CN X-ray counterparts"\ for short and use the term "X-ray emission from CN counterparts"\ to denote the X-ray emission detected from historical CNe in quiescence or in the post-nova supersoft X-ray phase. To search for and study such counterparts, we use data from three SRG/eROSITA sky surveys in the hemisphere analyzed by the RU eROSITA   consortium is responsible. Our goal is (i) to study the X-ray properties of a large sample of CN counterparts, (ii) to search for counterparts in the post-nova supersoft X-ray phase, and (iii) to construct the accretion rate distribution of CN counterparts in a binary system and to compare it with the one for known steady supersoft sources and with theoretical predictions.

\section{Sample of Historical Classical Novae in the Galaxy}
%----------------------------------------------------------------------------------------------
\begin{figure}[t]
\hspace{1mm}{\includegraphics[width=\linewidth]{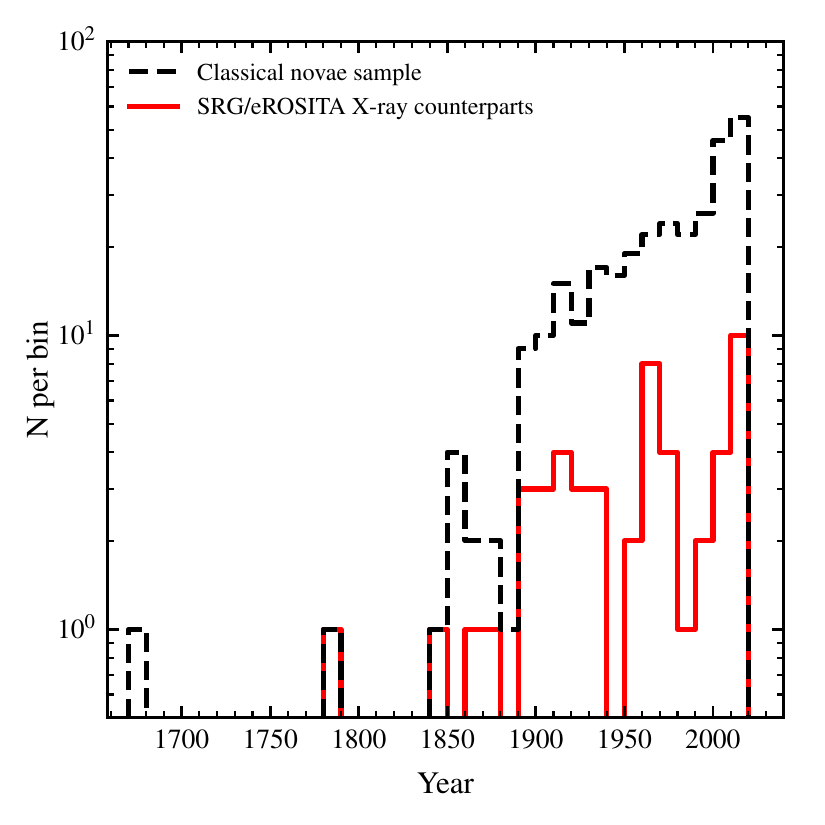}}
\caption{ Distribution of Galactic CNe in the half of the sky being processed by the Russian SRG/eROSITA consortium in years of their detection: the dashed (black) line indicates the complete CN sample; the solid (red) line indicates the CHs having cross-matches with X-ray sources from the SRG/eROSITA catalog within 15$\arcsec$.}
\label{fig:Nova_year_distribution}
\end{figure}
%-----------------------------------------------------------------------------------

To study the X-ray emission from CN counterparts in quiescence, we collected a sample of known historical CNe in our Galaxy that were classified and confirmed based on their optical light curves and spectra. The list of known CNe compiled by the Central Bureau for Astronomical Telegrams\footnote{\url{http://www.cbat.eps.harvard.edu/nova\_list.html}} was taken as a basis. This list includes the CNe discovered from 1612 to 2010. We added the CNe detected from 2010 to 2021 from the publicly accessible lists of Koji Mukai\footnote{\url{https://asd.gsfc.nasa.gov/Koji.Mukai/novae/novae.html}} and Bill Gray\footnote{\url{https://projectpluto.com/galnovae/galnovae.htm}} to this list. We selected only those counterparts that are located in the half of the sky for the processing of which the Russian SRG/eROSITA consortium is responsible into the final CN sample.

The resulting sample of CNe in our Galaxy produced in this way includes 309 sources with Galactic longitudes
$0\degr \le l\le 180\degr$ discovered from 1670 to 2021. The distribution of CNe in years of their detection is presented in Fig. \ref{fig:Nova_year_distribution}. As can be seen from the figure, most of the events were detected after 1850.

\section{SRG/\lowercase{e}ROSITA Telescope Data}

In this paper, we use the data from three sky surveys obtained by the SRG/eROSITA telescope from December 2019 to June 2021. The eROSITA data were preprocessed and calibrated at the Space Research Institute of the Russian Academy of Sciences by the eROSITA data processing system developed and maintained by the X-ray catalog science working group of the Russian eROSITA consortium using some elements of the eSASS (eROSITA Science Analysis Software) package developed at the Max Planck Institute for Extraterrestrial physics. In the data processing, we used the results of ground-based preflight calibrations and flight calibration observations in October–November 2019 and during 2020–2021. The data from the three sky surveys were combined to increase the sensitivity.

The X-ray catalog of sources was cross-correlated with the list of known CNe in the Galaxy with a search radius of $\rm 15\arcsec$. At such a search radius we found 52 matches. At the same time, the expected number of chance matches is $\sim 0.75$, i.e., fairly small. The final list of X-ray sources having identifications with CNe is presented in Table \ref{tab:nova_list}. The CN distances and color excesses were taken from various papers the references to which are given in the table. We determined the distances for some of the CNe using the measured parallaxes from the second data release of the Gaia catalog of sources \citep{2016A&A...595A...1G,2018A&A...616A...1G}. To cross-match the CN list and the Gaia catalog, we used a search radius of 1$\arcsec$. The Gaia data were used only for three sources for which the proper motions and/or parallaxes were measured with a statistical significance $\ga 3\sigma$.

When extracting the spectra and light curves, a circle with a radius of $60\arcsec$ was used as the source region, and an annulus with inner and outer radii of $120\arcsec$ and $240\arcsec$, respectively, was used as the background region. X-ray spectra were analyzed using XSPEC v.12 software \citep{1996ASPC..101...17A}. The spectrum was approximated using the $C$-statistics \citep{1979ApJ...228..939C}. In this case, the spectral channels were grouped in such a way as to include at least three counts per channel\footnote{see note for work in XSPEC:\\ {https://heasarc.gsfc.nasa.gov/xanadu/xspec/manual/ \\ XSappendixStatistics.html}}.

To test the quality of the fit to the data by the spectral model, we used Monte Carlo simulations. In particular, we used the realization of such simulations provided in the XSPEC package by the {\it goodness} command with the Anderson–Darling (AD) statistical test\footnote{see Appendix B in XSPEC manual}. The test using the {\it goodness} command allows the percentage of the simulations with the AD statistic greater than the actual value obtained when fitting the observed source spectrum by the chosen model to be calculated. For each source and each spectral model, we produced $\rm 10^{5}$ realizations with the
{\it nosim} and {\it fit} parameters. In each realization, the spectrum was drawn according to the Poisson statistic using the best-fit model parameters and then was fitted by the same model to determine the best value of the AD statistic. The distribution of the derived values of the AD statistic was used to estimate the probability to obtain a value of the statistic exceeding the observed one as a result of statistical fluctuations. To compare the fits to the same data by different spectral models, we used the Akaike information criterion (AIC) \citep{1974ITAC...19..716A} calculated as as $AIC=2\times k+C$, where $k$ is the number of free parameters in the fit, and $C$ is the value of $C$-statistics for the best-fit parameters. The preferred model for fitting the observed spectrum is the one with the minimum value of $AIC$.

\section{X-ray Emission from the Counterparts of Classical Novae in Quiescence}

%----------------------------------------------------------------------------------------------
\begin{figure}[t]
\hspace{1mm}{\includegraphics[width=\linewidth]{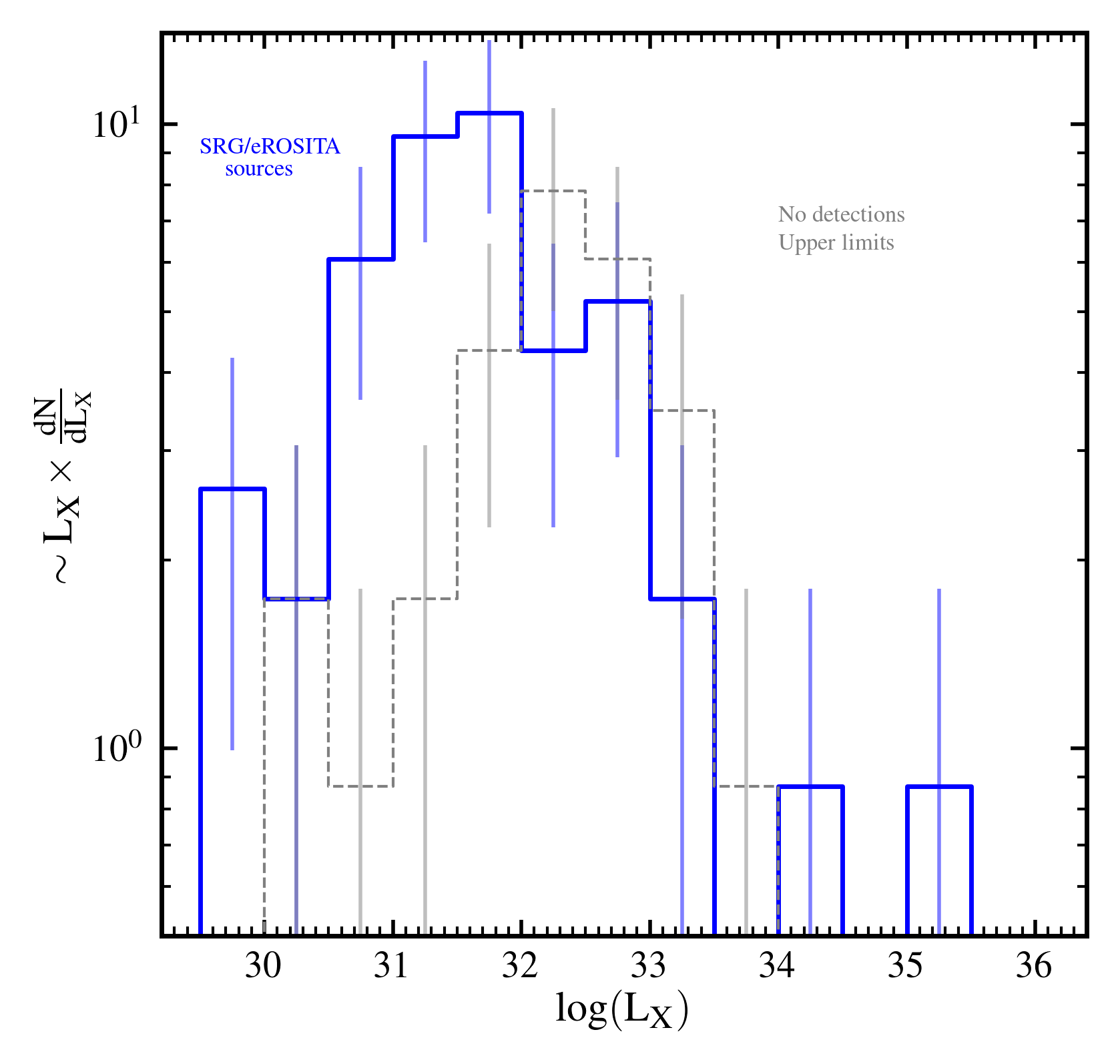}}
\caption{Distribution of CN counterparts in observed X-ray 0.3--2.3 keV luminosity from the data of the three SRG/eROSITA sky surveys (blue color). The gray histogram indicates the distribution of 3$\sigma$ upper limits on the X-ray luminosities of the CN counterparts that were not detected based on the data of the three sky surveys for which the distances are known. The X-ray luminosity was corrected for interstellar absorption. The observed distributions uncorrected for the sample incompleteness effects are shown. }
\label{fig:Lx_distr}
\end{figure}
%-----------------------------------------------------------------------------------

\subsection{Spectral Analysis of Bright Sources}

For 11 sources from our sample, the SRG/eROSITA telescope recorded more than 50 counts, which is enough for a spectral analysis. Their spectra are shown in Fig. \ref{fig:brights_spectra}, while their fits by various spectral models are given in Table \ref{tab:spectra_list}. To take into account the interstellar absorption, we used the Tubingen-Boulder ISM absorption model ({\it tbabs} model in XSPEC, \citealt{2000ApJ...542..914W}) under the assumption of solar elemental abundances in the interstellar medium.

The spectra of the bright sources exhibit some dichotomy between the sources with hard spectra, $\Gamma\sim 1$, and the sources with soft spectra, $\Gamma\sim 2-3$, which is discussed in more detail in this and succeeding
sections. We found no obvious correlation of the spectral shape with the source luminosity or with the time elapsed since the CN outburst.

The spectra of several sources have a more complex shape than the simple power-law model or the optically thin plasma model. Therefore, we also used the isobaric cooling flow model {\it mkcflow}\footnote{Note that when using the {\it mkcflow} model, it is necessary to set the redshift parameter $\rm z>0$, since the model was originally created to describe X-ray emission from a cluster of galaxies. We calculated the redshifts from the distances given in Table \ref{tab:nova_list} and the Hubble constant $H_0=69.3$ km/(Mpc s).} in XSPEC \citep{1988ASIC..229...53M} to describe the observed spectra. This model describes successfully the X-ray spectra of dwarf novae in quiescence \citep[see, e.g.,][]{2003ApJ...586L..77M,2005MNRAS.357..626B,2005ApJ...626..396P,2009ApJ...707..652M,
2017PASJ...69...10W}. The minimum and maximum plasma temperatures were free parameters in spectral fitting. If the best-fit value of the minimum plasma temperature is made coincident with zero, then this parameter will be fixed at 8 eV. For a number of sources from our sample, in particular, for V603 Aql, V2487 Oph, V2491 Cyg, and VY Aqr, the {\it mkcflow} model provides a better fit to the observed X-ray spectra than does the simple power-law model or, in some cases, the single-temperature optically thin thermal plasma model. At the same time, for the sources with hard spectra GK Per, V392 Per, X Ser, and V4743 Sgr, for which the power-law fit gives a photon index $\Gamma \sim 1$, the isobaric cooling flow model is clearly inapplicable and leads to unreasonably high temperatures $\ga 100$ keV unrealizable on the WD surface. For these sources, we do not provide the parameters of the fit by the {\it mkcflow} model.

The spectrum of V2491 Cyg defies a satisfactory description by any of the simple single-component models considered above. This source has a comparatively soft spectrum, but the optically thin single temperature
plasma model does not describe the emission from the source. Adding another component, black body radiation, to the model improves significantly the quality of the spectral fit, making it acceptable. In the case of such a two-component model, the data do not limit the temperature of the optically thin thermal component from above and, therefore, it was fixed at 10 keV. The nature of the supersoft component in the spectrum of V2491 Cyg is discussed below.

The spectrum of V2487 Oph also defies description by the spectrum of an optically thin thermal plasma, but the isobaric cooling flow model provides a satisfactory quality of the fit to the data. However, just as in the case of V2491 Cyg, the possible presence of a supersoft component in the spectrum of V2487 Oph was noted in previous studies \citep[see, e.g.,][]{2002Sci...298..393H}. Therefore, we fitted the spectrum of this source by a two-component model including black body radiation and achieved a significant improvement in the quality of the fit to the data compared to the simple single-temperature thermal model. The parameters, the temperature and luminosity of the soft component $kT\approx100$ eV and $\log(L_X)\approx36.3$, are comparable to those for V2491 Cyg. However, a comparison of the C-statistic and the Akaike information criterion shows that the {\it mkcflow} model is sufficient to fit the source spectrum, and the introduction of a second component is not required. Thus, the question of whether a supersoft component is present in the spectrum of V2487 Oph remains open. For other sources from our sample, adding a supersoft thermal component does not improve the quality of the spectral fit.

In the spectra of some sources, in particular, V603 Aql, V2487 Oph, and V392 Per, the spectral features that can be interpreted as emission or absorption lines engage our attention. In the case of V603 Aql and V2487 Oph, these features located at energies $\approx 3.8$ and $\approx 4.3$ keV do not correspond to emission lines of any cosmically abundant elements. Taking into account the degrees of freedom associated with the positions of these lines and their widths, their statistical significance is low, and they do not require a separate discussion. In the case of V392 Per, the absorption line is located at an energy $\approx 0.68\pm 0.02$ keV, which is close to the resonance OVII and OVIII lines. Assuming that the position of the line is fixed, its statistical significance is $\approx 2-2.5\sigma$ for one studied spectrum. Taking into account the number of studied spectra, this is quite a moderate significance and, therefore, we do not consider this as a reliable detection. Note that no absorption lines of ionized oxygen have been observed previously in the spectrum of V392 Per.

\subsection{Average Spectrum of Faint Sources}

%----------------------------------------------------------------------------------------------
\begin{figure}[t]
\hspace{1mm}{\includegraphics[width=\linewidth]{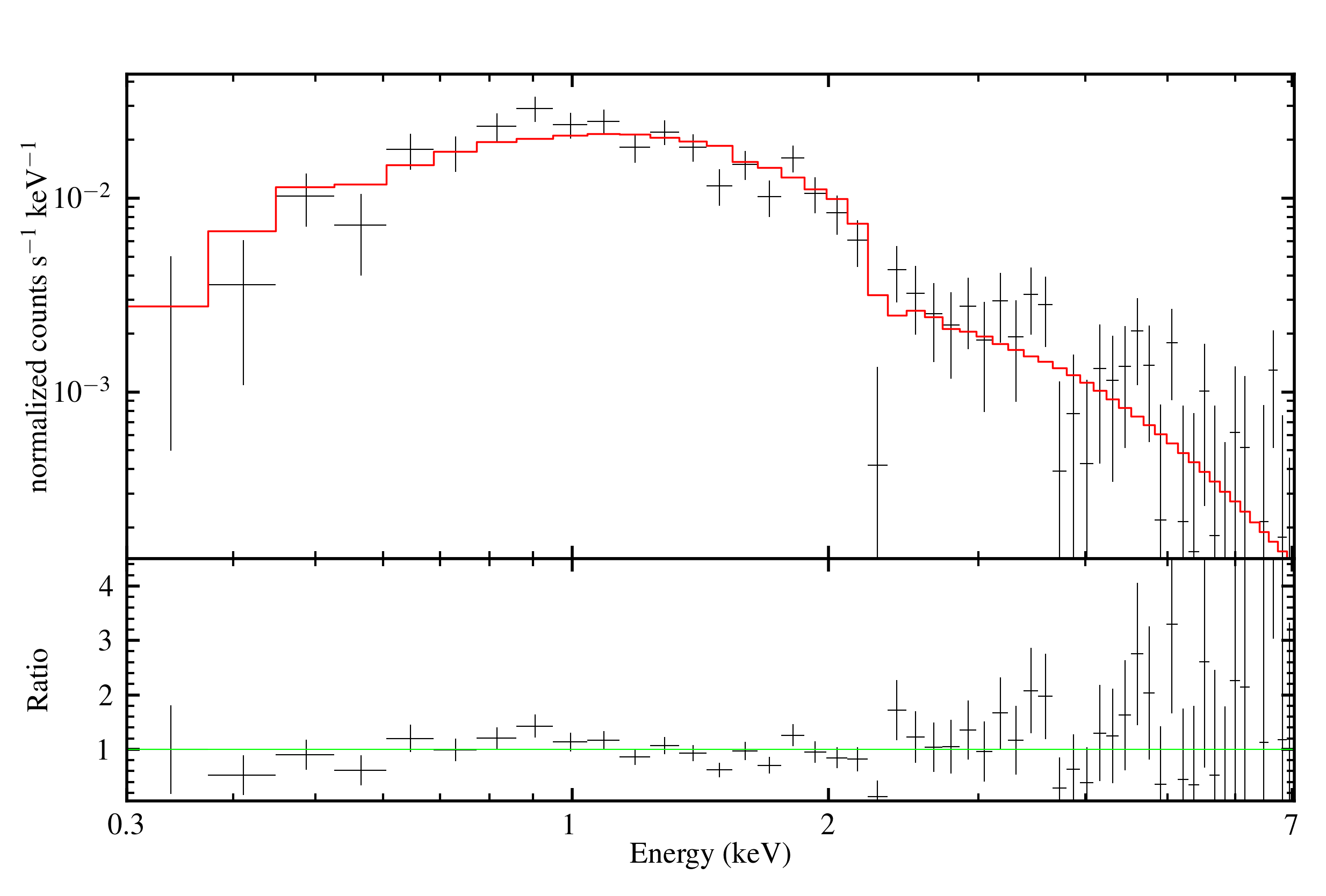}}
\caption{Averaged X-ray spectrum of the sources with fewer than 50 counts. The model of a power-law slope is represented by the red color. Lower panel: the ratio of the observed counts to the model ones in each energy channel.}
\label{fig:comb_spectra}
\end{figure}
%-----------------------------------------------------------------------------------

We constructed the average spectrum of the sources from which fewer than 50 counts were recorded and whose signal-to-noise ratio was not enough for a detailed individual study. The averaged spectrum of these sources is shown in Fig. \ref{fig:comb_spectra}. Fitting the spectrum by a power law gives the best-fit parameters $\Gamma=0.96^{+0.18}_{-0.17}$ and $N_H=9.9^{+3.8}_{-3.4}\times10^{20}$ cm$^{-2}$. As expected, based on the spectral hardness, fitting the spectrum by the optically thin plasma model does not allow an upper limit for the plasma temperature to be obtained. The lower limit on the temperature is $\rm \approx 43$ keV. We also divided the faint sources into two groups: the sources from which from 25 to 50 counts were recorded (12 sources) and the sources with fewer than 25 counts (25 sources). The average spectra for both groups have parameters compatible, within the error limits, with those given above.

\subsection{X-ray Luminosities}

To calculate the X-ray luminosities of the sources from which, on average, more than 50 counts are recorded, we used the model with the minimum value of AIC, corresponding to the best-fit model. For the fainter sources, we used the spectral model constructed by fitting the average spectrum of the faint sources presented above.

We calculated the hydrogen column density for each source individually based on measurements of the color excess $E(B-V)$. To convert the color excess to the extinction $\rm A_V$, we assumed $R_{V}\approx3.1$ \citep{1989ApJ...345..245C}, while we used the standard formula $N_{H}\approx 2.21 \times 10^{21}\times A_{V}$ \citep{2009MNRAS.400.2050G} to calculate the column density. For those sources where the color excess $E(B-V)$ was not known, we used the total hydrogen column density in the Galaxy toward the source derived from HI4PI data \citep{2016A&A...594A.116H}. Since the typical spectra of CN counterparts are fairly hard (except for several sources discussed in the next section), the part of the bolometric correction associated with the correction of the X-ray luminosity for low-frequency absorption is small. Therefore, some uncertainty in $N_{H}$ for the sources with an unknown color excess does not affect strongly our results. Some uncertainty in the spectral shape of the faint sources does not affect the accuracy of their X-ray luminosity estimates either. For example, variations in the slope of the expected power-law spectrum in the range 0.9–2.7 lead to a variation in the counts-to-physical fluxes conversion factor by $\rm \pm{15}\%$.

The distances are known for 41 of the 52 sources, which allows their luminosities to be calculated. For the remaining 11 sources, we estimated their luminosities using the median distance of the sources with known distances, which turned out to be 1.5 kpc. Thus, we derived the distribution of the detected sources in their X-ray 0.3–2.3 keV luminosity corrected for absorption. The derived distribution is shown in Fig. \ref{fig:Lx_distr}. The remaining 257 historical CNe from our sample were not detected in the data of the
three SRG/eROSITA sky surveys. For 31 of these sources, we determined the distances based on the Gaia catalog and constructed the observed distribution of 3$\sigma$ upper limits on their X-ray luminosities (see Fig. \ref{fig:Lx_distr}).

\section{Outburst of the Dwarf Nova VY Aqr}

In November 2020, an outburst of the dwarf nova VY Aqr occurred, which reached its maximum light on November 8, 2020, $\sim 9.6^m$ in the visual band \citep{2020AAN...724....1W}. By chance coincidence, the source was scanned on this day by the SRG/eROSITA telescope during the second sky survey, which allowed its behavior in X-rays to be studied at the maximum of the optical light curve. During the first (May 6, 2020) and third (May 9, 2021) sky surveys, the source was in quiescence.

The SRG/eROSITA data allow the X-ray spectrum of VY Aqr during the dwarf nova outburst to be analyzed for the first time. To study the spectrum in quiescence, we combined the spectra of the first and third sky surveys to increase the statistics. The spectral fits by various models are given in the last two columns of Table \ref{tab:spectra_list}.

The power-law model slope and the optically thin plasma temperature show that during the dwarf nova outburst the source spectrum becomes softer, whereas its luminosity increases by a factor of $\rm \approx10$. The characteristic temperatures in the isobaric cooling flow model in outburst decrease by a factor of $\approx2$. During a dwarf nova outburst, an increase in the accretion rate gives rise to an optically thick boundary layer, which leads to an apparent “softening” of the X-ray spectrum \citep[e.g.,][]{1977MNRAS.178..195P,1985ApJ...292..550P}. Note that in November 2007 VY Aqr was observed by the Suzaku observatory in quiescence \citep{2017PASJ...69...10W}. To fit the spectra, \citet{2017PASJ...69...10W} also used the cooling flow model to obtain a maximum temperature $kT_{max}\approx18.4$ and an accretion rate $\dot{M}_{acc}\approx 7\times 10^{-13}\ M_\odot$/yr, close to our values.

\section{Candidates for supersoft X-ray Source among the SRG/\lowercase{e}ROSITA Sample of CN Counterparts}

%------------------------------------------------------------------------
%--------------------------------------------------------------------------
\begin{figure*}
    \centering
     \includegraphics[width=0.47\linewidth]{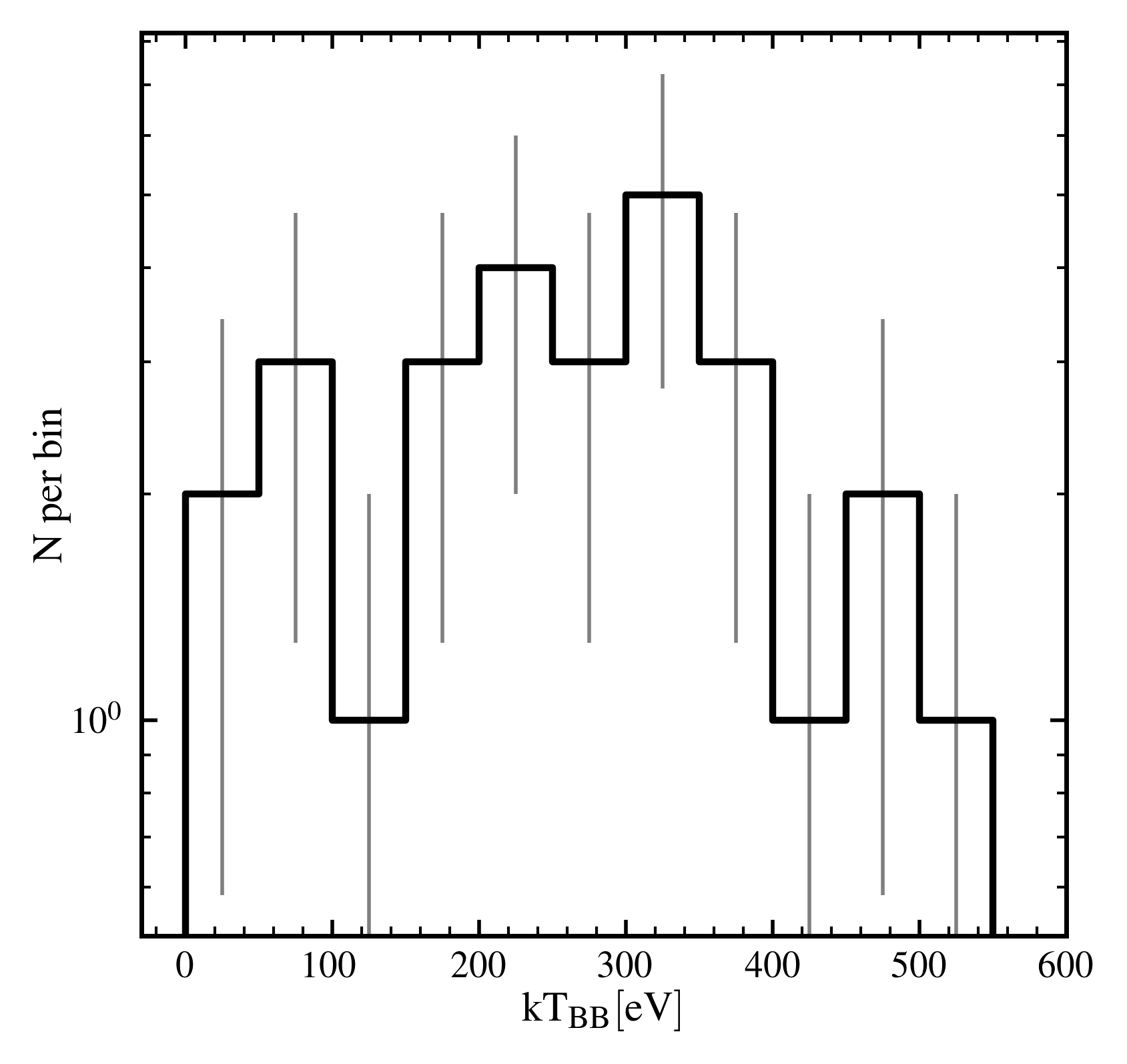}
    \includegraphics[width=0.45\linewidth]{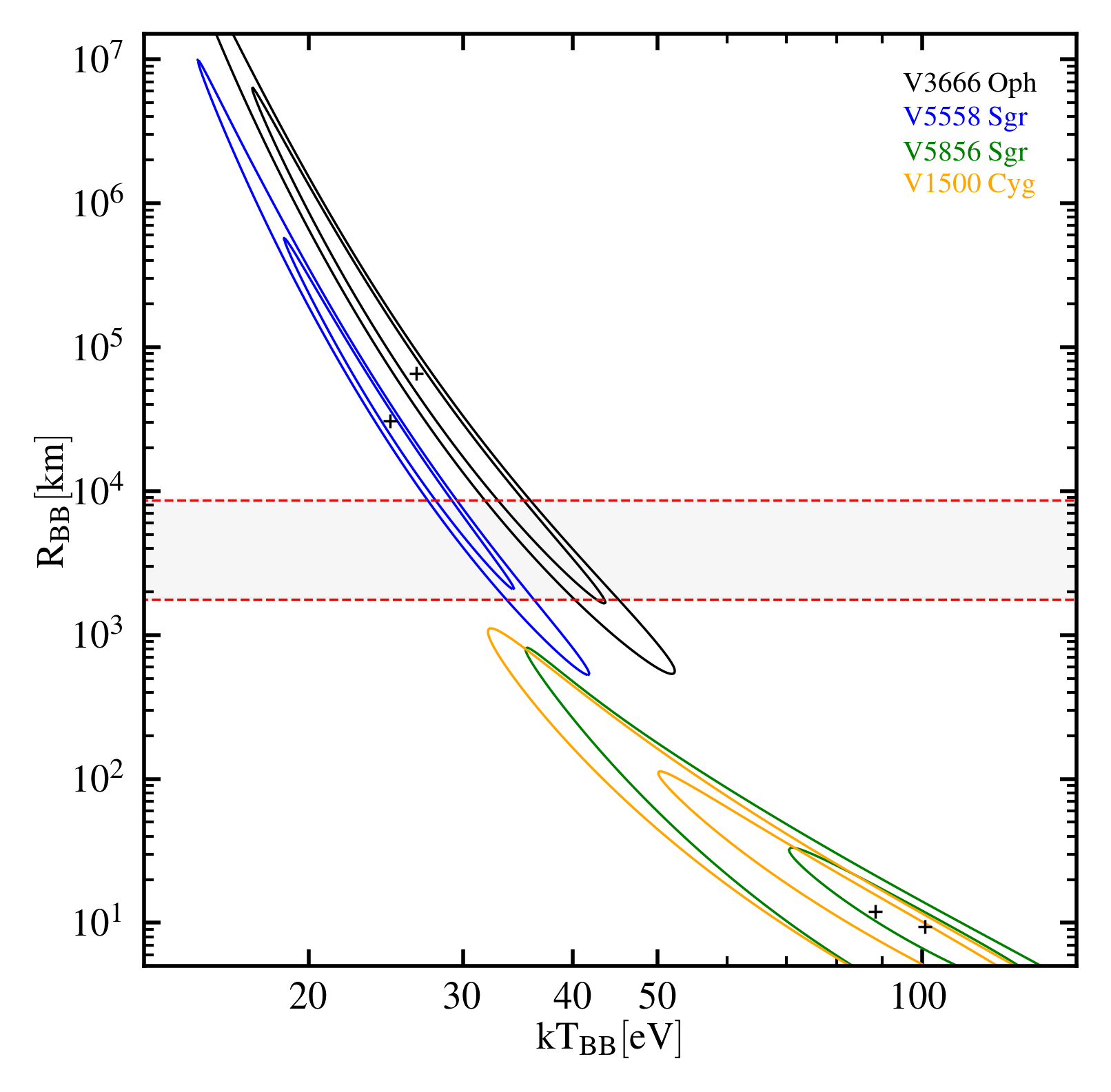}
    \caption{(a) Black body temperature distribution of the CN counterparts. The sources with the temperatures $\rm \ge 600$ eV are not included. The candidates for supersoft X-ray sources were preselected by temperature $\rm \le 200$ eV. (b): The contours of the 68\% and 95\% confidence intervals for the radius and temperature of the black body model. The red lines shows the WD radii for 0.6 $M_\odot$ and 1.4 $\rm M_\odot$, obtained from the mass–radius relation for zero-temperature WDs from \citet{2000A&A...353..970P}.
    }
    \label{fig:post_novae_SSSs}
\end{figure*}
%--------------------------------------------------------------------------

%------------------------------------------------------------------------
\begin{figure*}
    \centering
    \includegraphics[width=45mm]{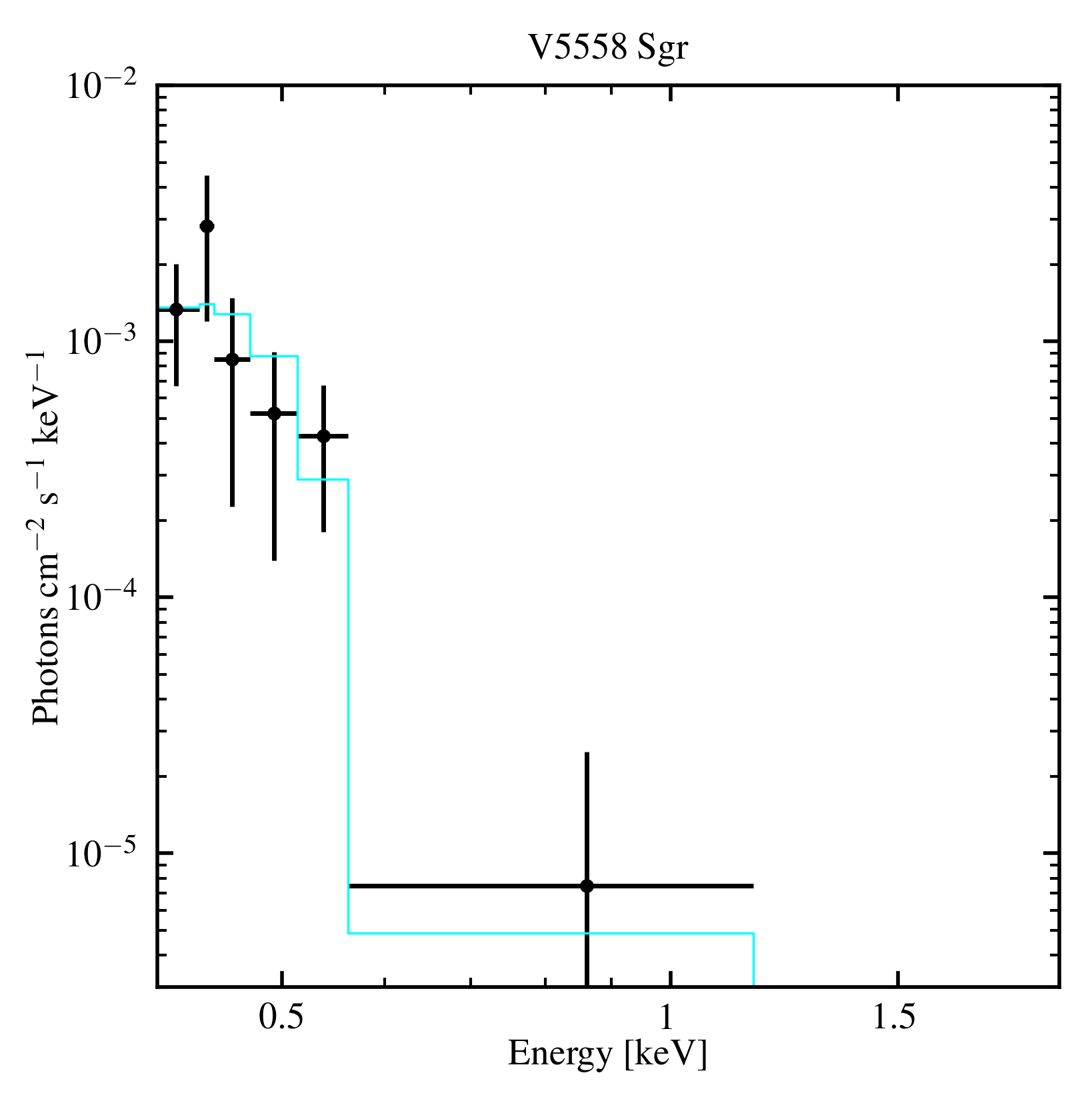}
      \includegraphics[width=45mm]{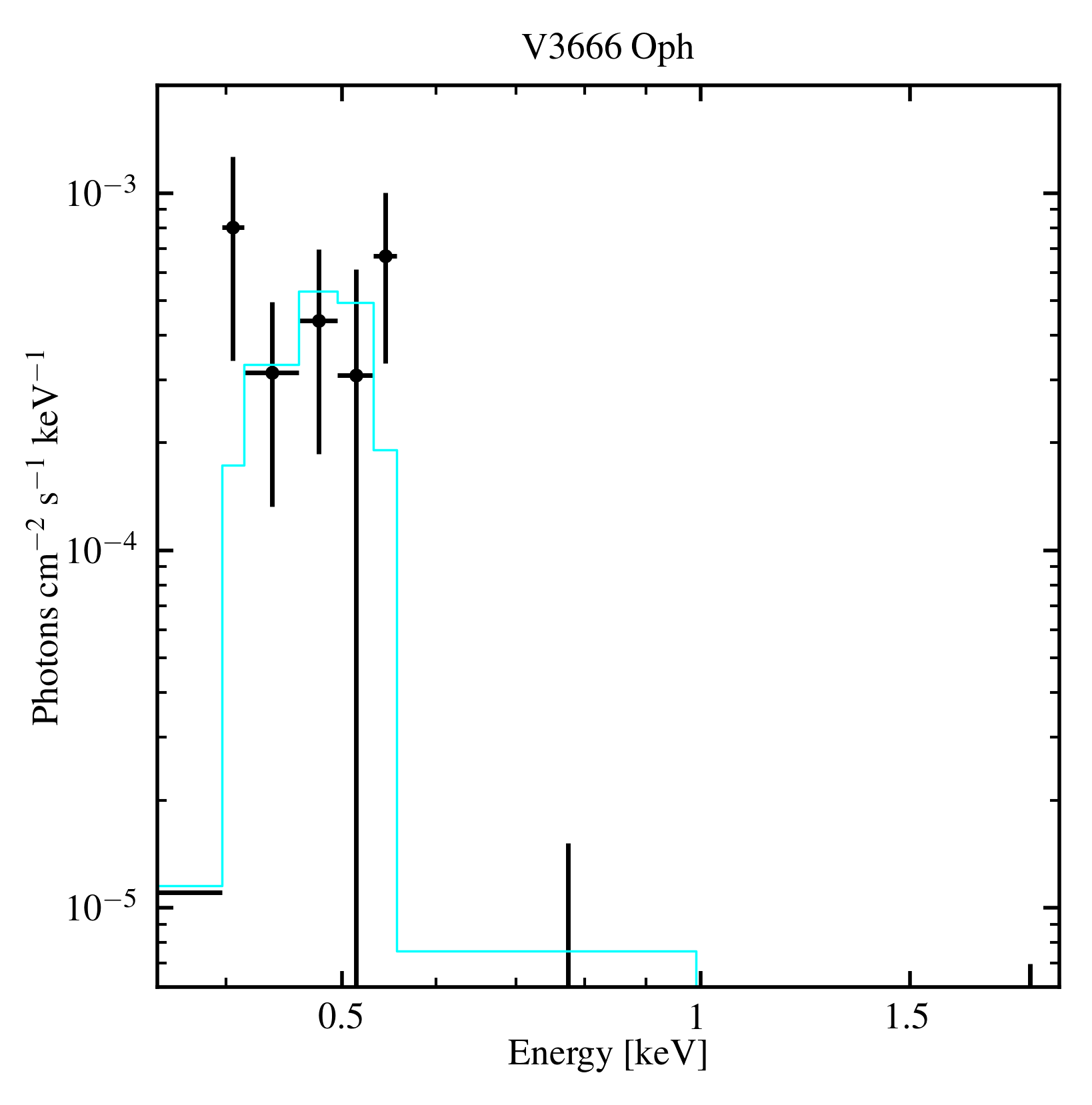} \\
    \includegraphics[width=45mm]{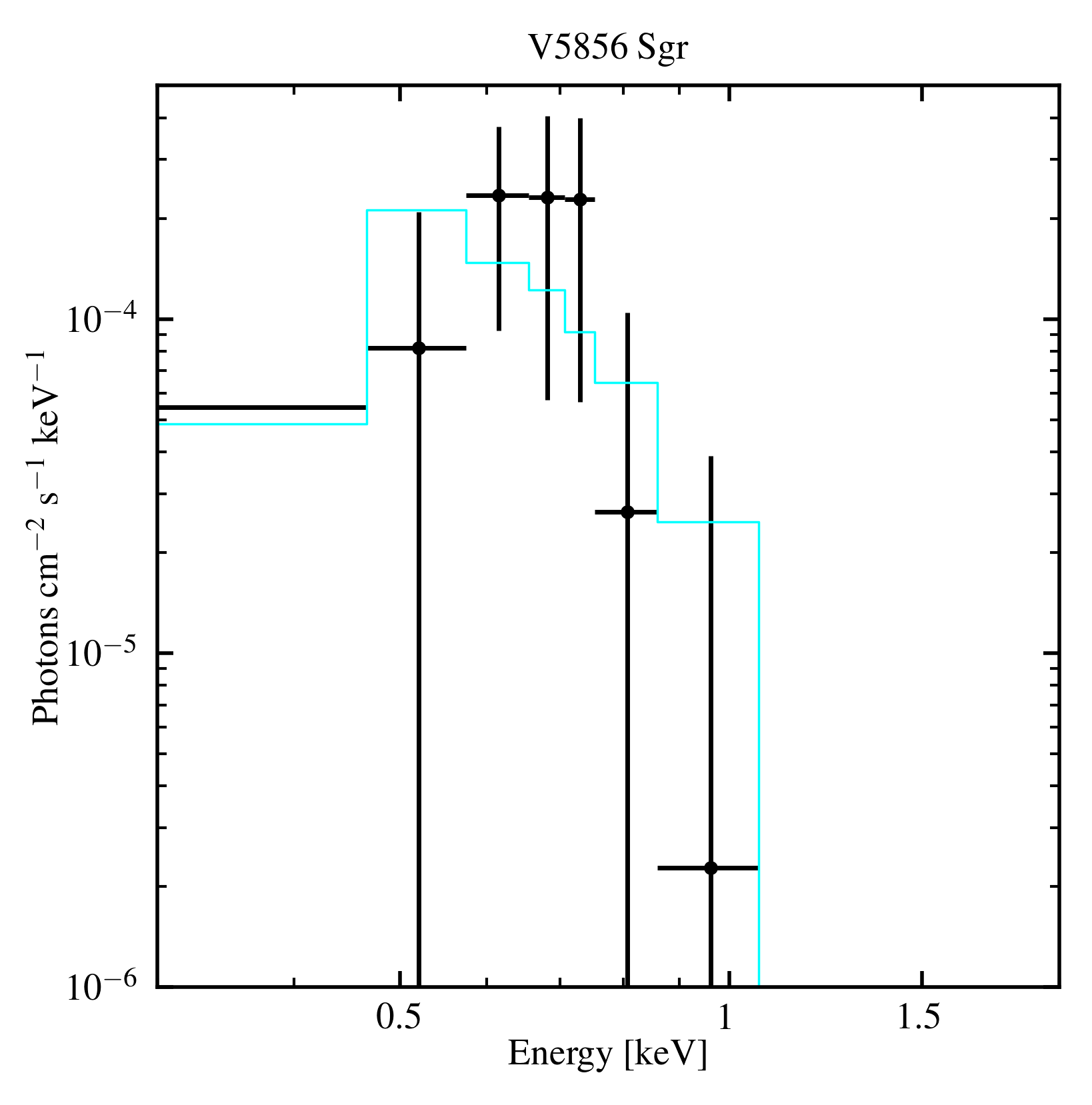}
    \includegraphics[width=45mm]{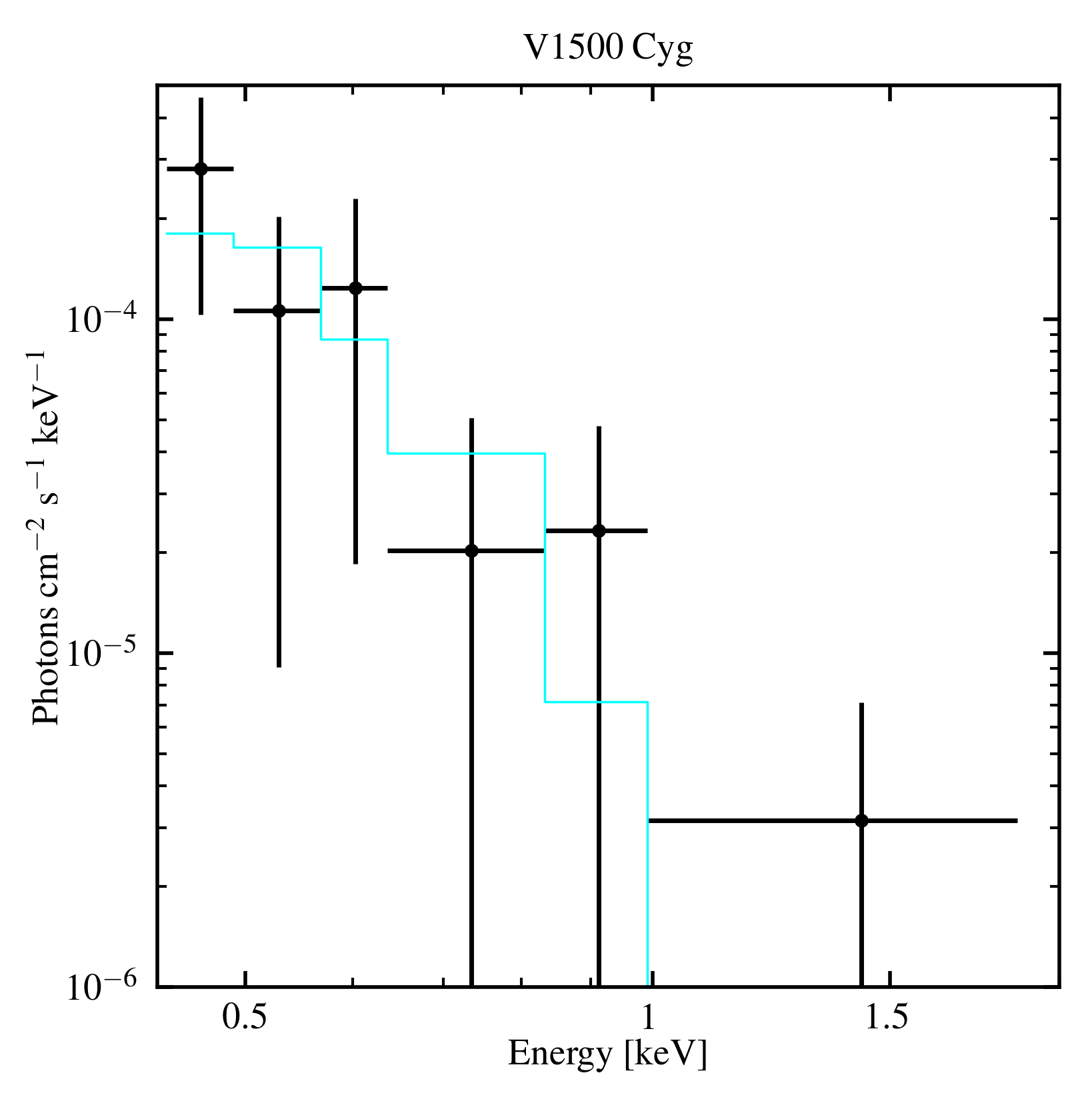}
    \caption{Spectra of potential candidates for supersoft X-ray sources among the CNe sample obtained with SRG/eROSITA. The black body models are shown in blue in each panel.}
    \label{fig:SSSs_spectra}
\end{figure*}

Among the CN X-ray counterparts, we searched for candidates for supersoft sources whose X-ray emission is associated with the post-nova supersoft X-ray phase. For this purpose, we fitted the spectra of all sources by a black body model with absorption (${\it tbabs}\times{\it bbodyrad}$ model in XSPEC). The hydrogen column density was fixed at the value determined from the color excess $E(B-V)$, as described above.

Most of the sources in our sample have rather hard X-ray spectra and are not described by the black body model. The approximation of such spectra by this model leads to high temperatures $\ga 200$ eV and low X-ray luminosities $L_X\approx 10^{30}\sim 10^{34}$ erg/s, which are inconsistent with the values expected during the post-nova phase. However, this procedure makes it possible to select sources with soft spectra for further investigation. The temperature distribution of sources is shown in Fig. \ref{fig:post_novae_SSSs}a, where we have excluded sources with temperature $\ga 600$ eV.

For further investigation, we selected nine sources with temperatures $\le 200$ eV. Five of them had a statistically significant ($\rm \ge 3\sigma$) flux above 1.0 keV and, after visual inspection, they were excluded from the sample of supersoft source candidates. As a result, out of 52 CNe sources, we selected four potential supersoft X-ray sources candidates (see their spectra in Fig. \ref{fig:SSSs_spectra}). The Fig. \ref{fig:post_novae_SSSs}b shows the confidence interval contours for the radius and temperature of black body model. We also shows the range of WD radii for the masses of 0.6 $M_\odot$  and 1.4 $M_\odot$ from the mass-radius relation for zero-temperature WDs \citep{2000A&A...353..970P}. Below we discuss these four supersoft source candidates in more detail.

{\bf SRGe J174224.2-205309 (V3666 Oph).} The approximation of the source spectrum by a black body model at a fixed column density $N_H\approx6.6\times 10^{21}$ cm$\rm^{-2}$ gives a temperature of $kT\approx 27\pm10$ eV and the $C$-statistics 6.86 for seven degrees of freedom. The observed X-ray luminosity in the 0.3--2.3 keV energy band is $L_{\rm X,obs}=1.5\pm0.5\times10^{31}$ erg/s. Due to the low temperature and the relatively high hydrogen column density, the bolometric correction of the luminosity is $\sim 10^6$ and the bolometric luminosity (the radius of the emitting surface) is highly uncertain: $L_{\rm bol}\approx 2.1^{+25.7}_{-2.0} \times 10^{37}$ erg/s. Since the distance to the source was unknown, a median distance of 1.5 kpc was used to estimate the luminosity. The confidence interval for the radius of the emitting surface of the black body model overlaps with the WD region; therefore, the X-ray emission from the source can be associated with the post-nova phase. Supersoft X-ray emission from the source is detected $\approx 1.6$ years after the CN outburst.

{\bf SRGe J181018.2-184653 (V5558 Sgr).} Similarly, the approximation of source spectrum by the black body model at a fixed column density  $N_H\approx5.5\times 10^{21}$ cm$\rm^{-2}$ gives a temperature of $kT\approx 25^{+10}_{-7}$ eV (the $C$-statistic value is 2.8 for five degrees of freedom). The observed X-ray luminosity in the 0.3--2.3 keV energy band is $L_{\rm X,obs}=2.8\pm0.6\times10^{31}$ erg/s, and the bolometric luminosity is $L_{\rm bol}\approx 1.8^{+28.8}_{-1.7} \times 10^{37}$ erg/s. The confidence region for the radius and temperature of the black body model includes the range of WD radii, therefore, the X-ray emission from the source might be associated with the post-nova phase. The supersoft X-ray emission from the source is detected after $\approx 13$ years of the CN outburst.

{\bf SRGe J182052.9-282218 (V5856 Sgr).} Similar to other two sources, we approximated the spectra of this source with a black body model with a fixed column density $N_H\approx2.3\times 10^{21}$ cm$\rm^{-2}$, and get a temperature of $kT\approx 101^{+51}_{-28}$ eV (the $C$-statistic value is 3.1 for five degrees of freedom). The observed X-ray luminosity  is $L_{\rm X,obs}=2.9\pm0.6 \times10^{32}$ erg/s in 0.3–2.3 keV energy band, and the bolometric luminosity is $L_{\rm bol}\approx 1.1^{+3.6}_{-0.8} \times 10^{33}$ erg/s. The confidence region  for the radius and temperature in the black body model does not overlap with the range of WD radii, therefore, the X-ray emission from the source is unlikely to be associated with post-nova phase. The supersoft X-ray emission from the source is detected after $\approx 3.4$ years the CN outburst.

We also note, that the parameters for the soft X-ray emission from V5856 Sgr are almost identical to those obtained for V1500 Cyg polar (see Fig. \ref{fig:post_novae_SSSs}b). However, this is not enough to classify V5856 Sgr as a polar. This hypothesis deserves a more detailed study in the course of future observations.

{\bf SRGe J211136.5+480905 (V1500 Cyg).} V1500 Cyg is a well-known polar in our Galaxy that shows a multicomponent X-ray spectrum. The soft component is described by a black body model with a temperature of $kT\approx60$ eV, while the hard component is described by bremsstrahlung model with a temperature of $kT_{br}\approx40$ keV \citep[e.g.,][]{2016MNRAS.459.4161H}. Due to relatively short exposure time during the sky survey, the hard component is not detected in its spectrum, and V1500 Cyg is observed by SRG/eROSITA as a supersoft X-ray source.

For this source the black body model approximation at a fixed column density, $N_H\approx3.1\times 10^{21}$ cm$\rm^{-2}$, gives a temperature of $kT\approx 88^{+36}_{-35}$ eV (the $C$-statistic value is 2.5 for six degrees of freedom). The observed X-ray luminosity is $L_{\rm X,obs}=2.9\pm0.6\times10^{31}$ erg/s in 0.3–2.3 keV energy band, and bolometric luminosity is $L_{\rm bol}\approx5.1^{+3.6}_{-2.9} \times 10^{32}$ erg/s.

\section{Observed Accretion Rate Distribution of Sources}

The historical CNe represent a sample of confirmed WDs with non-steady hydrogen burning on their surface. Measuring their post-outburst X-ray luminosity in quiescence allows the accretion rate in these systems to be estimated or limited from above. On the other hand, the steady supersoft sources observed in a bright state with a luminosity $\ga 10^{37}$ erg/s over tens of years represent examples of the sources with steady (or nearly steady) nuclear hydrogen burning. The accretion rate in such sources is easy to estimate by assuming that their luminosity is due to thermonuclear hydrogen burning reactions. A comparison of the derived accretion rate distributions for these two types of accretion WDs allows the predictions of the theory of thermonuclear burning on the WD surface to be tested.

%------------------------------------------------------------------------
\begin{figure*}
    \centering
    \includegraphics[width=0.47\linewidth]{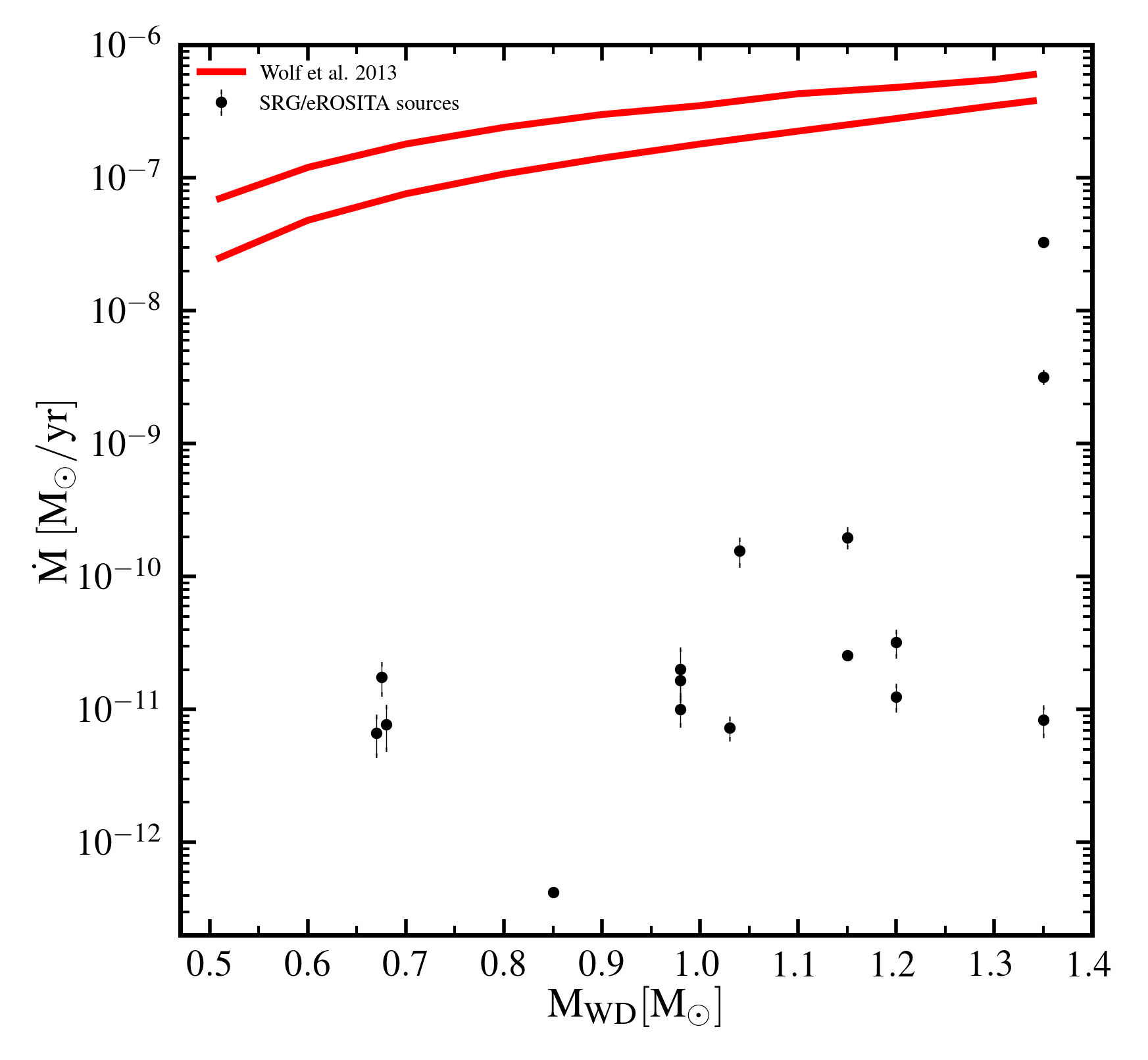}
    \includegraphics[width=0.45\linewidth]{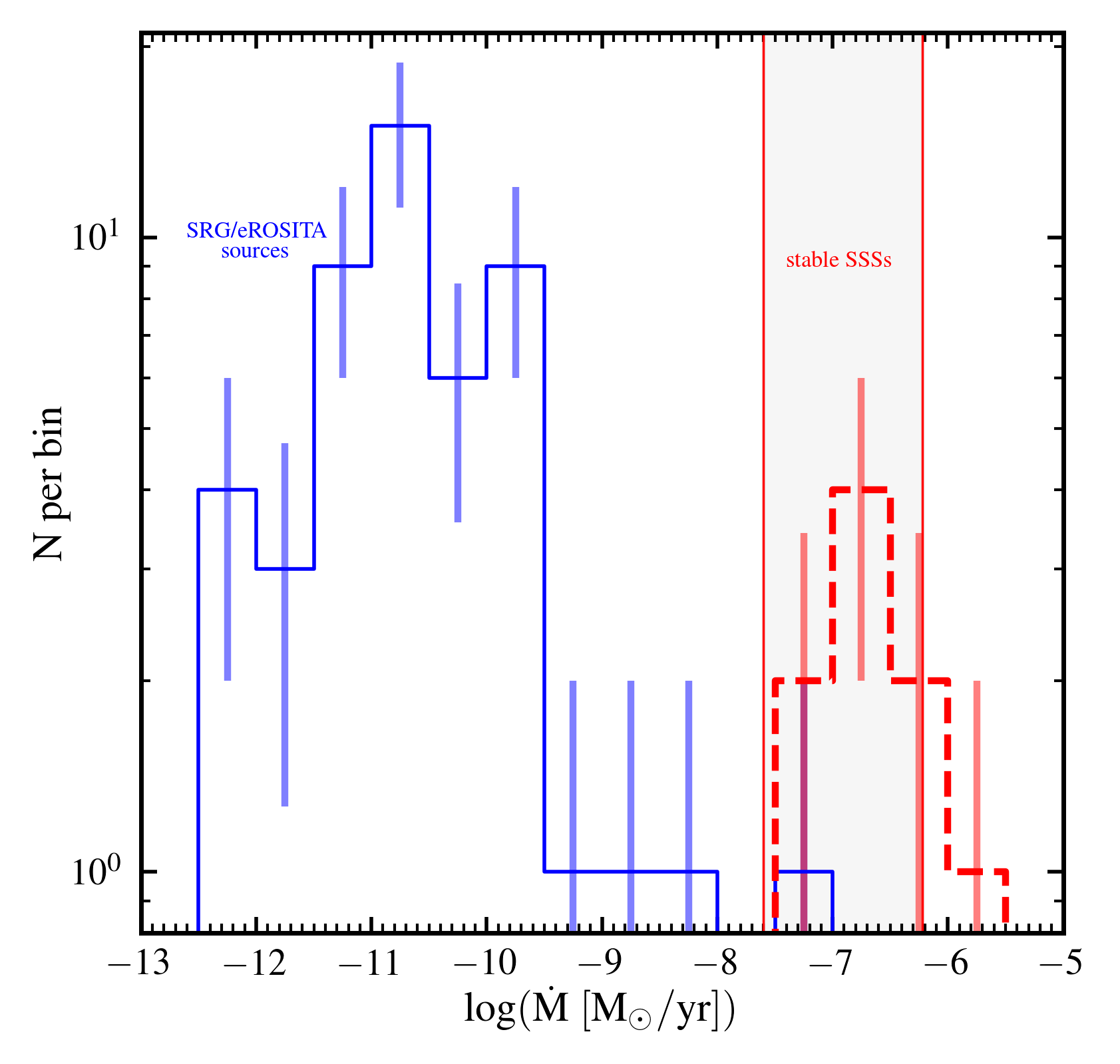}
    \caption{
   (a) Observed distribution of CNe with known WD masses on the $M_{\rm WD}-\dot{M}_{\rm acc}$ plane. The region of steady thermonuclear burning on the surface of an accreting WD obtained from the calculations by \citet{2013ApJ...777..136W} is represented by the (solid red lines). (b) Observed accretion rate distribution of all CN X-ray counterparts (blue color) and known stable supersoft X-ray sources (red color, dashed lines) from the catalog by \citet{1996LNP...472..299G}. The right panel is a projection of the $M_{\rm WD}-\dot{M}_{\rm acc}$ plane. The vertical red lines on the right panel are the boundaries of the stability strip for $\dot{M}_{acc}\approx 2.5\times 10^{-8}\ M_\odot$/yr  $(M_{WD}\approx0.51$ $\rm M_\odot$) and $\dot{M}_{acc}\approx 6\times 10^{-7}\ M_\odot$/yr  $(M_{WD}\approx1.34$ $\rm M_\odot)$. The CN counterpart falling inside the steady accretion zone in the right panel is  2487 Oph. As it can be seen from the left hand panel, this is a projection affect due to the large mass of the WD in this system. In the $M_{\rm WD}-\dot{M}_{\rm acc}$ plane this source is located well below the stability strip.
    }
    \label{fig:Mdot_panels}
\end{figure*}
%------------------------------------------------------------------------

\subsection{Steady Supersoft X-ray Sources}

In the case of thermonuclear hydrogen burning on the surface of an accreting WD, the bolometric luminosity of the source is defined as
\begin{equation}
\L_{\rm bol}=\epsilon_H\ X_H\ \dot{M}_{\rm acc},
\label{eq:H_burn}
\end{equation}
where  $\epsilon_H\approx6\times 10^{18}$ erg/g is the energy release in hydrogen fusion per unit mass of hydrogen, $X_H\approx0.72$ is the hydrogen mass fraction of accreting material and $\rm \dot{M}_{acc}$ is the mass accretion rate. Using this equation, we calculated the accretion rates for 10 known stationary supersoft X-ray sources from the \citet{1996LNP...472..299G} catalogue.

\subsection{CN Counterparts in Quiescence}

Most of the CN counterparts are in quiescence, except for several supersoft X-ray sources associated with recent CNe and the source VY Aqr that was observed by SRG/eROSITA at the peak of the dwarf nova outburst. These sources were discussed in detail in previous sections. In quiescence, during the accretion of material onto the WD, half of the radiated energy is released in the accretion disk and half is released in the boundary layer near the WD surface. At the accretion rates under discussion, $\dot{M}_{acc}\la 10^{-9}\ M_\odot$/yr, the boundary layer is optically thin \citep[see, e.g.,][]{1985ApJ...292..535P}, while the accretion disk is optically thick \citep[][]{1988AdSpR...8b.135S}. Some part of the accretion energy is converted into radiation in the optically thin hot corona of the accretion disk \citep[see, e.g.,][]{1984A&A...132..143M,1994A&A...288..175M}. The radiation from the accretion disk proper has a characteristic temperature $kT\la10 $ eV and is mainly outside the SRG/eROSITA energy range, while the observed X-ray emission is produced by the hot corona of the disk and the boundary layer. Hence, the accretion rate and the X-ray luminosity are related by the relation
\begin{equation}
 L_X=\eta \times \frac{G\ M_{*}\ \dot{M}_{acc}}{R_{*}},
\label{eq:Lacc_Mdot}
\end{equation}
where by the X-ray luminosity we mean the luminosity in the wide energy range 0.1–100 keV , $\rm \eta$ is the fraction of energy emitted in X-ray energy band ($\rm \eta \la 1$), $\dot{M}_{\rm acc}$ is the accretion rate, $M_*$ and $R_*$ - are mass and radius of the WD. 

The parameter $\eta$ depends on the ratio of the fractions of the energy radiated in different parts of the accretion stream—in the accretion disk, the disk corona, and the boundary layer. It follows from the virial theorem that in the case of a slowly rotating WD the fraction of the energy liberated in the boundary
layer is 50\%. Therefore, given that the boundary layer is optically thin, we expect $\eta\ga 0.5$. The contribution of the radiation from the accretion disk corona will increase $\eta$, but, obviously, the condition $\eta<1$ must be fulfilled. In the calculations below, we assumed $\eta=0.5$.

To estimate the bolometric correction for the X-ray luminosity measured in the 0.3--2.3 keV energy band, we assumed the source spectrum to be described by a bremsstrahlung model. As the bremsstrahlung temperature
varies in the range 5--20 keV, the bolometric correction varies in the range $L_X/L_{0.3-2.3}\approx 2.4-5.4$. For the subsequent calculations, we assumed $kT_{br}=10$ keV and the bolometric correction to be $\approx 3.5$

In those cases where theWD mass was unknown, we assumed that it was  $M_{WD}=0.8$ $\rm M_\odot$. The WD radius was calculated from the mass–radius relation by \citet{2000A&A...353..970P}.

\subsection{Accretion Rate Distribution of WDs}

Figure \ref{fig:Mdot_panels}a shows the CN counterparts for which there are WD mass measurements (see Table \ref{tab:nova_list}) on the $M_{\rm WD}-\dot{M}_{\rm acc}$ plane. As follows from this figure, all of the CN counterparts are located well below the “stability strip”, as might be expected from the fact that the sample of historical CNe consists of accreting WDs about which it is known for sure that hydrogen burning on the WD surface is non-steady. We constructed the “stability strip” in Fig. \ref{fig:Mdot_panels} based on the
calculations by \citet{2013ApJ...777..136W}.

Figure \ref{fig:Mdot_panels}b presents the accretion rate distribution of all the CN counterparts detected by SRG/eROSITA and known steady supersoft sources from the catalog by \citet{1996LNP...472..299G}. The projection of the theoretical “stability strip” onto the accretion rate axis is also shown in this figure. As can be seen from the figure, the sources with non-steady hydrogen burning on the WD surface ($\dot{M}_{\rm acc}\approx10^{-12}\sim10^{-8}\ M_\odot$/yr) and the stable supersoft sources ($\dot{M}_{\rm acc}\approx10^{-7.5}\sim10^{-6}\ M_\odot$/yr) occupy different ranges nonintersecting in accretion rate. On the one-dimensional accretion rate distribution, one source, V2487 Oph, falls into the region occupied by the steady supersoft sources. However, as can be seen from Fig. \ref{fig:Mdot_panels}a, this is a projection effect, and on the $M_{\rm WD}-\dot{M}_{\rm acc}$ plane V2487 Oph ($M_{\rm WD}\approx  1.35\ M_\odot$ 
and $\dot{M}_{\rm acc}\approx9\times10^{-8}\ M_\odot$/yr) is located below the “stability strip”.

\section{Discussion}

\subsection{Nature of the X-ray Emission from CN Counterparts}

Several mechanisms for the generation of X-ray emission after the end of a CN outburst from a non-magnetized WD are known: (i) in the shock when the CN ejecta interact with the interstellar medium, with the stellar wind from the companion or the ejecta from previous CN events \citep[see, e.g.,][]{1977ApJ...213..492B,2008clno.book.....B}; (ii) the post-nova supersoft X-ray emission resulting from residual hydrogen burning on the WD surface (for a discussion, see below); (iii) the emission from the accretion disk, the corona, and the boundary layer in CN quiescence, including the outbursts of dwarf novae.

In the first case, the X-ray emission of an optically thin thermal plasma with a characteristic luminosity $L_X \la 10^{32}-10^{34}$ erg/s is observed on time scales $\sim$100 days after the outburst \citep[e.g.,][]{2014MNRAS.442..713M}. Since there are no such recent CNe in our sample, this mechanism may be excluded from further consideration.

The post-nova supersoft X-ray emission also lasts for a relatively short time after the CN outburst (but considerably longer than the shock emission) and is easily identified by its supersoft spectrum. There are two such sources in our sample, and they are discussed below. Note that supersoft X-ray emission can also be observed in the case of magnetized WDs, polars. In this case, however, steady hydrogen burning occurs more often, and CN outbursts are observed more rarely. Interestingly, there is one confirmed polar and one more source with similar X-ray properties in our sample (for a discussion, see below).

The only mechanism for the generation of X-ray emission long after the CN outburst is the accretion of material from the donor star onto the WD. This mechanism explains the X-ray emission from most of the CN counterparts in our sample, as confirmed by the properties of their X-ray spectra, which allows the X-ray emission to be used as an indicator for measuring the accretion rate in these systems. As has already been discussed above, the boundary layer near the WD surface and the accretion disk corona make a major contribution to the X-ray emission.

Empirically, to describe the emission spectrum of the boundary layer in cataclysmic variables, the isobaric cooling flow model is successfully used. It describes gas cooling at constant pressure from a maximum temperature $kT_{\rm max}\sim 10-60$ keV to a minimum temperature $kT_{\rm min}\la 1$ keV \citep[e.g.,][]{2003ApJ...586L..77M,2005MNRAS.357..626B,2005ApJ...626..396P,2009ApJ...707..652M,2017PASJ...69...10W}. Some dichotomy is observed among the bright sources in our sample of CNe, for which we perform a detailed spectral analysis. While for most of the sources isobaric cooling flow model was preferable (power-law's photon index is typically $\Gamma\sim$ 2--3); for four sources - GK Per, V392 Per, X Ser, V4743 Sgr - this was not true. Approximation of their spectra by power-law model gives the photon index $\Gamma\sim 1$. The observed differences in the shape of the spectrum do not clearly correlate with the luminosity of the source. We note that one of the four sources with hard spectra, GK Per, is a confirmed intermediate polar, and V4743 Sgr is a candidate for intermediate polars \citep[][]{2016MNRAS.460.2744Z}, and the nature of other two sources is unknown, though they can be assumed as intermediate polars; however such a conclusion seems premature and requires more investigation.

The average spectrum of the faint sources (from which fewer than 50 counts were recorded in the survey) is also hard, with a photon index $\Gamma\sim 1$. Based on the available data, we cannot propose a justified interpretation of this result. However, note that the spectral hardness may suggest that an appreciable
fraction of these sources are (intermediate) polars with a low accretion rate. According to the catalog by \citet{2003A&A...404..301R}, among them there are one more confirmed intermediate polar (DQ Her) and two candidates (V533 Her and V2467 Cyg). Excluding these sources from the averaging does not change the shape of the average spectrum. Further, as SRG/eROSITA will scan whole sky more frequently, there will be enormous amount of data be collected and, the statistics for individual sources will almost triple, and a detailed spectral analysis of a larger number of sources will be possible, which will probably help clarify the mechanisms of X-ray emission in cataclysmic variables

\subsection{Steady and Non-steady Thermonuclear Hydrogen Burning on the Surface of an Accreting WD}

Depending on the accretion rate and the WD mass, hydrogen burning on the WD surface can be steady or non-steady. As follows from Fig. \ref{fig:Mdot_panels}, the steady supersoft X-ray sources are located, as
expected, in the “stability strip”. At the same time, all of the CN counterparts are located well below
the “stability strip”, in accordance with the predictions of the theory of thermonuclear burning on the WD surface.

We note that with the current selection method (which is based on the list of historical CNe), it is possible to obtain a sample of reliable accreting WDs with a non-steady thermonuclear burning of hydrogen on their surface. As far as we know, this is the first time such a sample design method has been used.

Additionally, we note that the existence of a “stability strip” is still a matter of debate. A number of calculations suggest that hydrogen thermonuclear burning on the WD surface is non-stationary in the entire range of the accretion rate and WD mass \citep[e.g.,][]{1995ApJ...445..789P,2005ApJ...623..398Y,2013IAUS..281..166S}. In these models, the amplitude of outbursts decreases with increasing accretion rate, while their frequency increases, with the mass loss decreasing, and such sources become close in their properties to (quasi-)steady supersoft sources. The SRG/eROSITA data under discussion do not allow us to distinguish between these two models.

\subsection{Post-Nova Supersoft X-ray Emission}

Among the 52 CN counterparts detected in Xrays, we found four supersoft X-ray sources. The luminosities and temperatures of two of them, V5558 Sgr and V3666 Oph, give a size of the emitting region compatible with the range of admissible WD sizes (note that for supersoft spectra there exists strong degeneracy between the temperature and luminosity or size). The interpretation of their supersoft X-ray emission as a result of residual hydrogen burning on the WD surface seems quite plausible.

The duration of the post-nova supersoft X-ray phase depends on the WD mass and the initial ejecta
mass \citep[see, e.g.,][]{2016MNRAS.455..668S}. Supersoft X-ray emission from V5558 Sgr and V3666 Oph is
detected $\rm \approx 13$ and $\rm \approx 1.6$ years after the CN outburst, respectively. The theoretical relation beyween WD mass and duration of post-nova phase allow us to estimate WD masses of these sources. Based on the theoretical light curves for the post-nova supersoft X-ray phase from \citet[][]{2016MNRAS.455..668S}, we expect that the WD mass in V5558 Sgr should not exceed $\la0.8 M_\odot$, while
the WD mass in V3666 Oph is $\la1\ M_\odot$. The above constraint on the WD mass in V5558 Sgr agrees well
with the mass that was found by analyzing the slope of the optical light curve and from optical spectroscopy,
$M_{\rm WD}\approx0.58-0.63\ M_\odot$ \citep{2010NewA...15..657P}.

To our knowledge, in our Galaxy there are three CNe with a post-nova supersoft X-ray phase duration of more than three years: V723 Cas (18--19 years)  \citep{2008AJ....135.1328N,2015ATel.8053....1N}, GQ Mus ($\approx$10 years) \citep{1995ApJ...438L..95S}, V574 Pup ($\approx$3.2 years) \citep{2011ApJS..197...31S}. V5558 Sgr might be the second longest post-nova in our Galaxy.

\subsection{Supersoft X-ray Emission from Polars and Their Candidates}

V1500 Cyg is a known polar in our Galaxy, and its spectrum exhibits a supersoft component, along with a hard X-ray emission. SRG/eROSITA detects only the supersoft component in the spectrum of this source, while the hard component is below the sensitivity threshold of the all-sky survey. The size of the emitting region of the supersoft component, $\la 104$ km (68\%), is much smaller than the WD size and corresponds to the size of the polar region at the magnetic WD pole \citep[][]{2003cvs..book.....W}. Such a supersoft component is often
observed in polars \citep[see, e.g.,][]{1996MNRAS.278..285R,2004MNRAS.347..497R}.

The supersoft X-ray emission from V5856 Sgr cannot be associated with the post-nova supersoft X-ray emission due to the small size of the emitting surface(see Fig. \ref{fig:post_novae_SSSs}). Interestingly, the parameters of the supersoft component in V5856 Sgr are identical to those in V1500 Cyg. Taking this into account, we can assume that V5856 Sgr is also a polar. Due to a small number of counts in the X-ray spectrum of V5856 Sgr, we do not detect the hard component, as in the case of V1500 Cyg. A more rigorous justification of this assumption requires further observations in the optical and X-ray spectral bands. We also note, that V5856 Sgr has been classified as one of the brightest sources in the $\gamma$--ray energy band, detected during the CN outburst \citep{2017NatAs...1..697L}.

The spectrum of V2491 Cyg is peculiar and it is the only source in our sample whose spectrum  cannot be described by any of the one-component models. Though V2491 Cyg was not selected as a supersoft X ray source in our search method, to describe its X-ray spectrum we require a soft black body component with the temperature of $\approx 65$ eV. At the time of SRG/eROSITA observations, 12 years elapsed since the CN outburst. The WD mass in this system is $M_{\rm WD}=1.35\ M_{\odot}$ (see Table \ref{tab:nova_list}). For such massive WDs, the duration of the post-nova phase does not exceed a $\sim$month \citep[][]{2016MNRAS.455..668S}. Moreover, the emitting surface area of the supersoft X-ray emission in this source is $\approx6\times10^{6}$ km$^2$, which is much less than the WD area $\approx 10^{8}$ km$^2$ for a WD with mass 1.35 $M_{\odot}$. These facts indicate that the observed supersoft component in the spectrum of V2491 Cyg cannot be associated with the X-ray post-nova phase.

Supersoft X-ray emission from this source was also detected in quiescence earlier. In particular, it was observed by the Suzaku $\rm \approx$ 2 years after the CN outburst. A supersoft component with
a temperature  $\rm \approx 77$ eV and a bolometric luminosity $\approx 1.4\times 10^{35}\times(d/10.5$ kpc) erg/s  was also detected in these observations \citep[see, e.g.,][]{2015ApJ...807...61Z}. Among the hypotheses under consideration, there is the assumption that this source is an intermediate polar. Unfortunately, the SRG/eROSITA data do not help to clarify the nature of this source.

\section{Conclusions}

In this paper, we studied the X-ray emission from historical CNe in our Galaxy. We used data from three SRG/eROSITA sky surveys in the half of the sky that is processed by the Russian SRG/eROSITA consortium. Our results can be summarized as follows:

(1) Out of the 309 known historical CNe  (see Fig. \ref{fig:Nova_year_distribution}, \ref{fig:Lx_distr}), X-ray emission was detected from 52 sources. Most of the sources are in quiescence — their X-ray emission is associated with the accretion of material in the binary system and is produced predominantly in the boundary layer near the WD surface and in the hot corona of the accretion disk. The X-ray 0.3–2.3 keV luminosities are $L_X\approx10^{30} - 10^{34}$ erg/s.

\begin{landscape}
\begin{figure*}
\begin{center}
\hbox
{
\includegraphics[width=60mm]{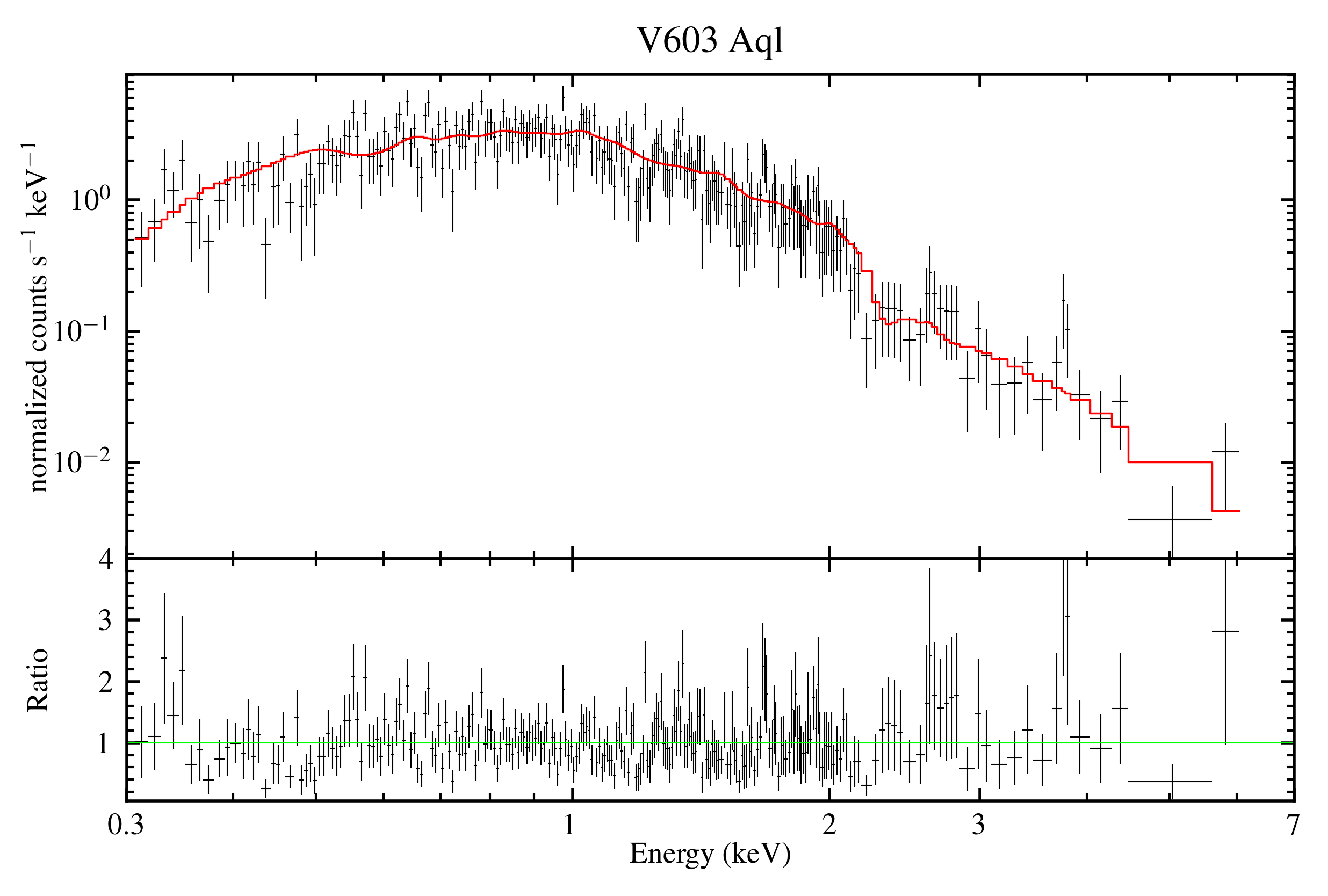}
\includegraphics[width=60mm]{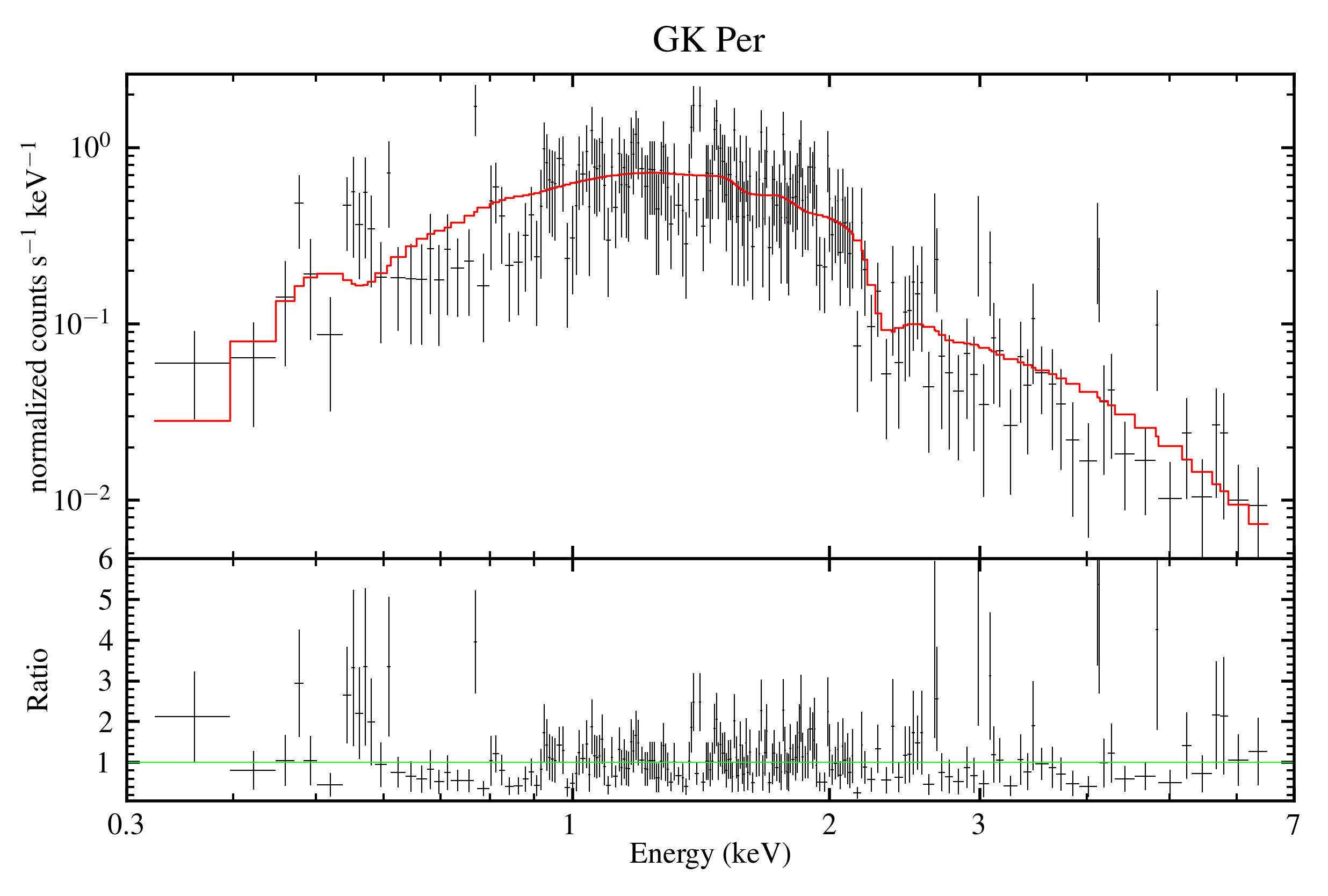}
\includegraphics[width=60mm]{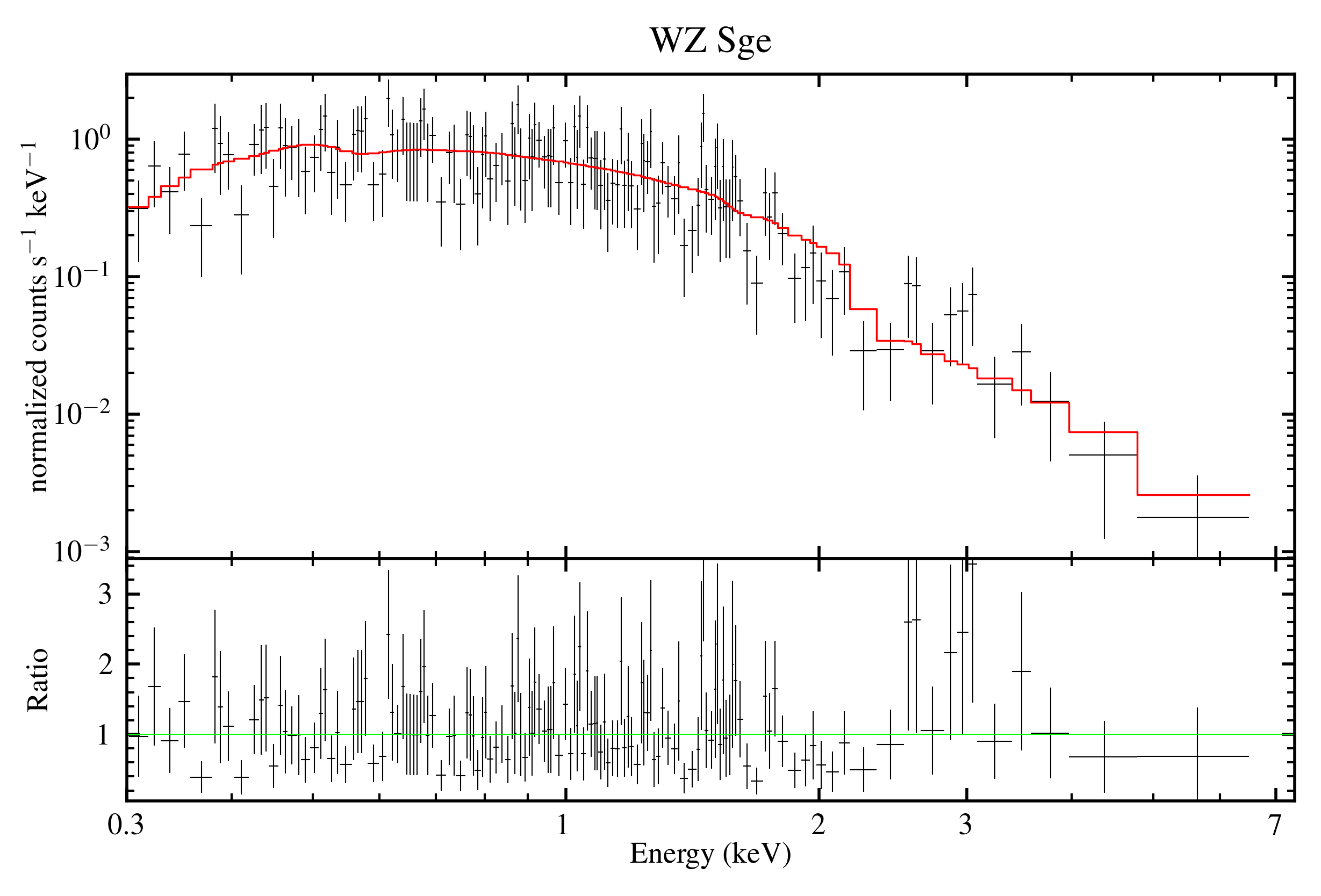}
\includegraphics[width=60mm]{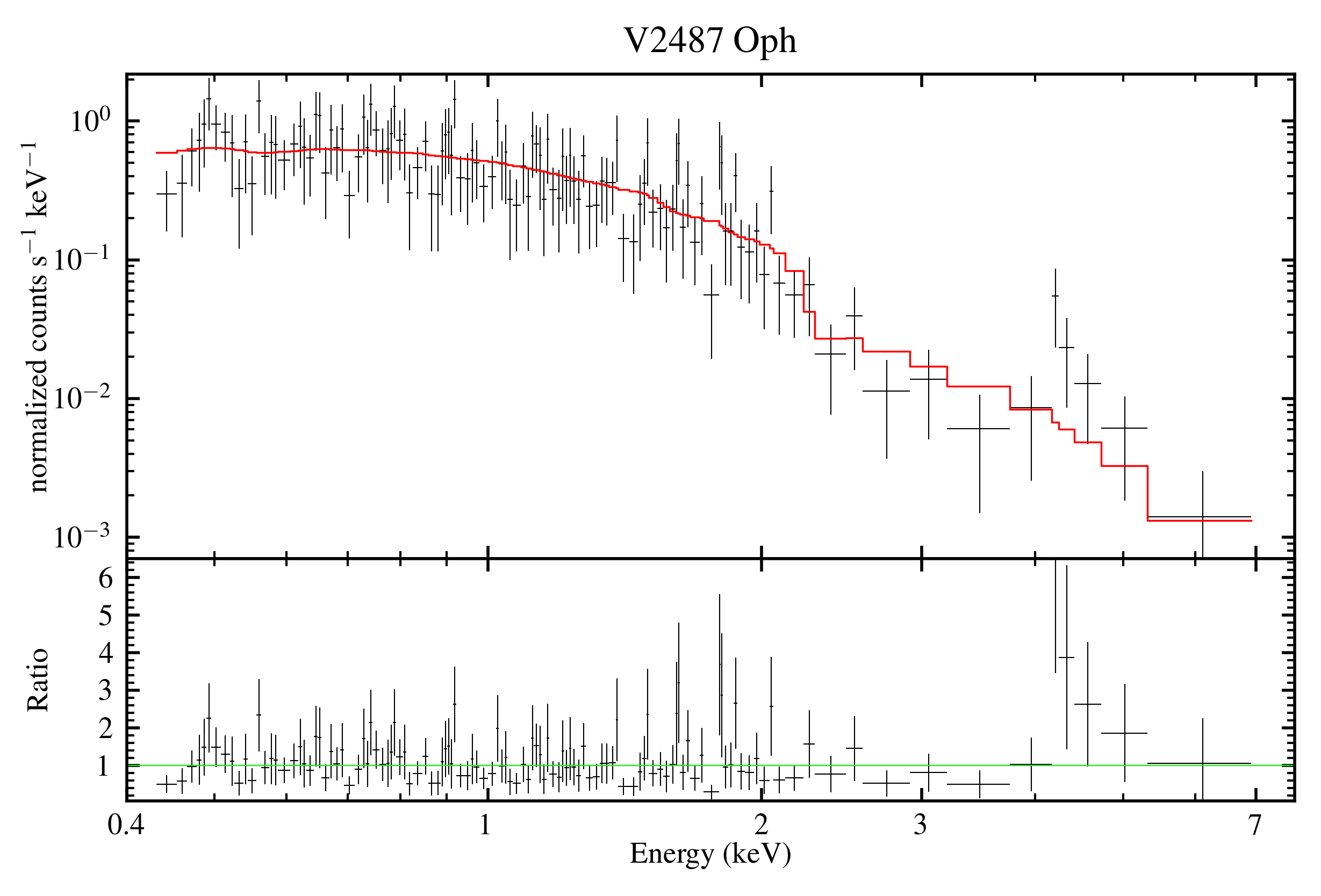}
}
\hbox
{
\includegraphics[width=60mm]{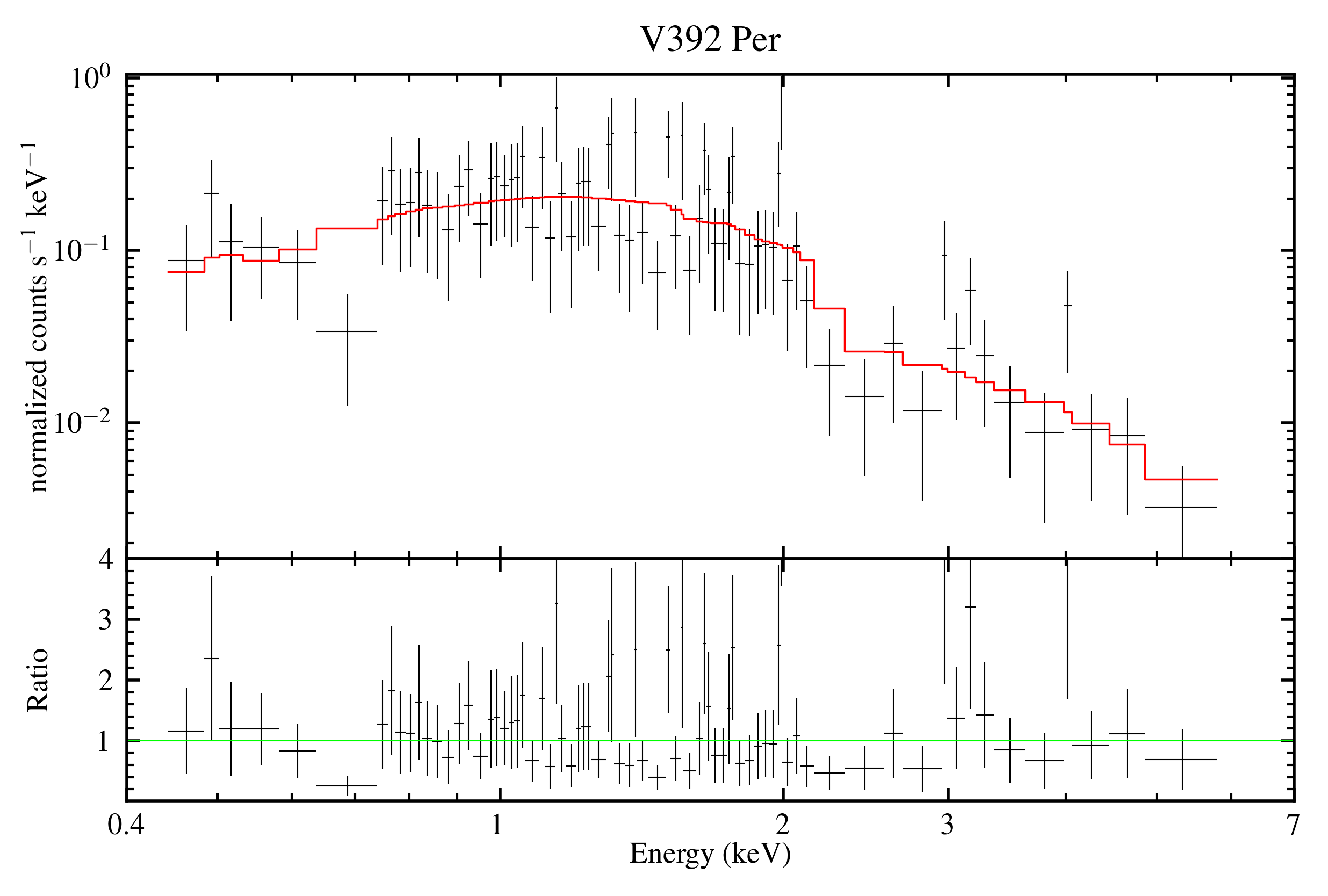}
\includegraphics[width=60mm]{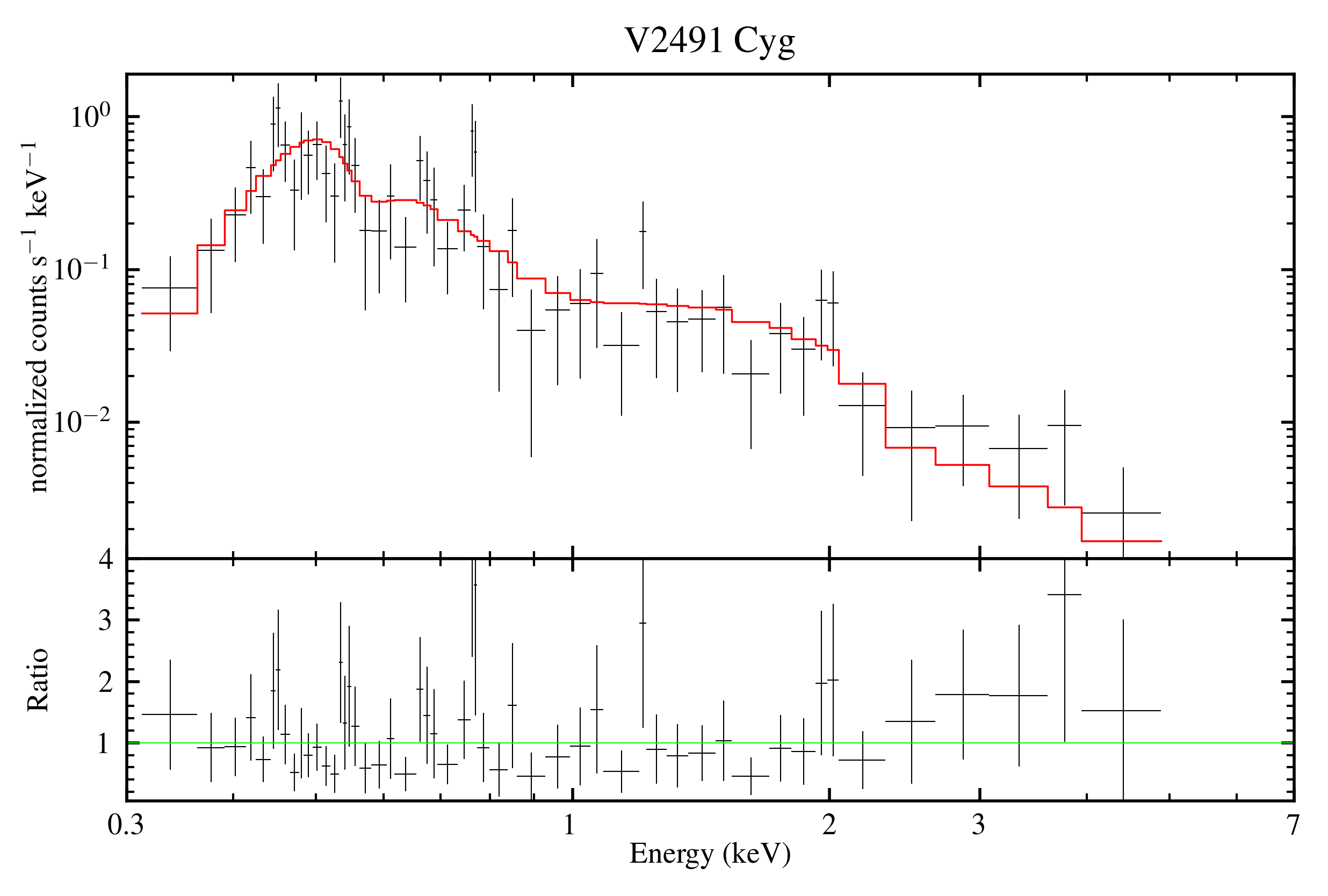}
\includegraphics[width=60mm]{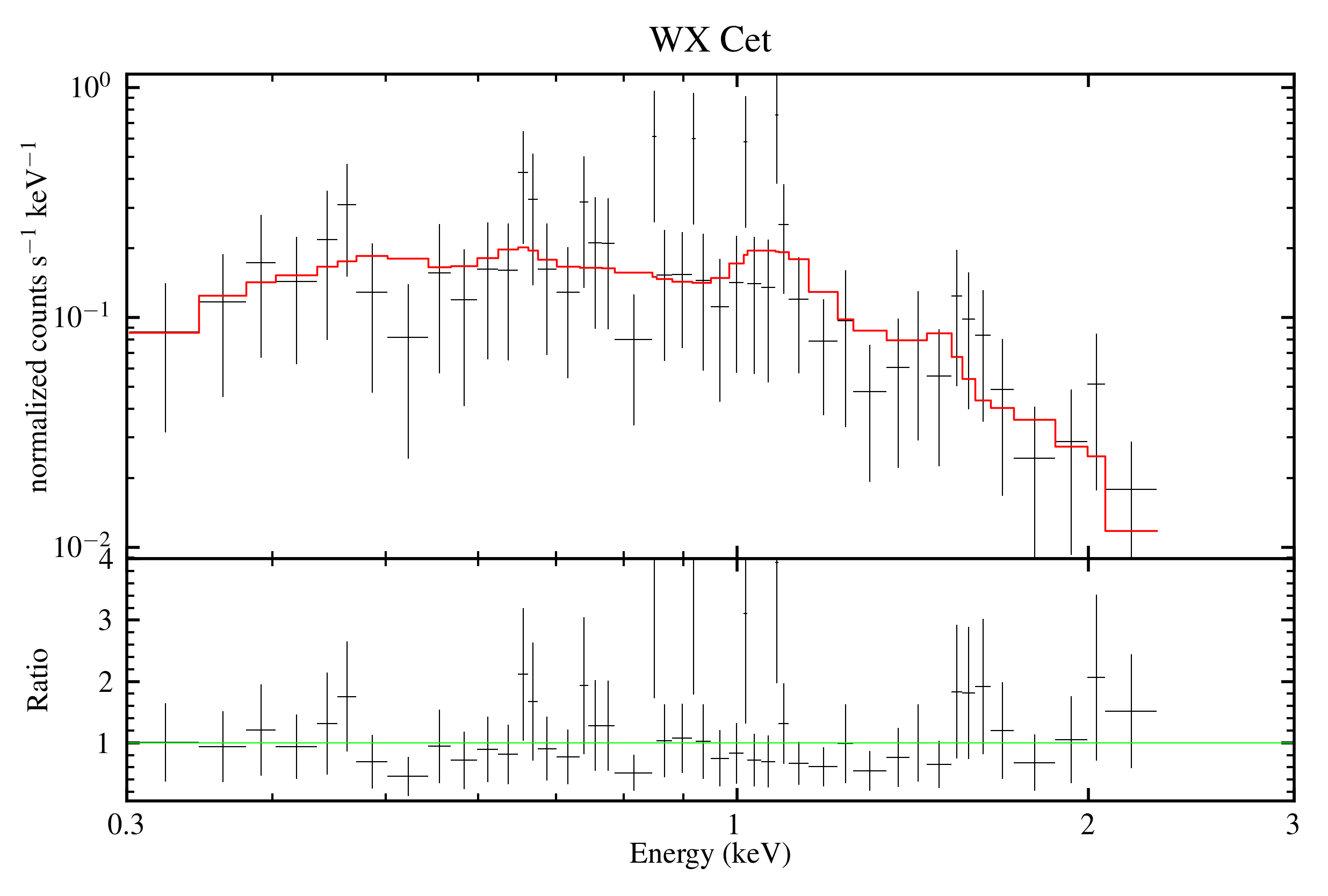}
\includegraphics[width=60mm]{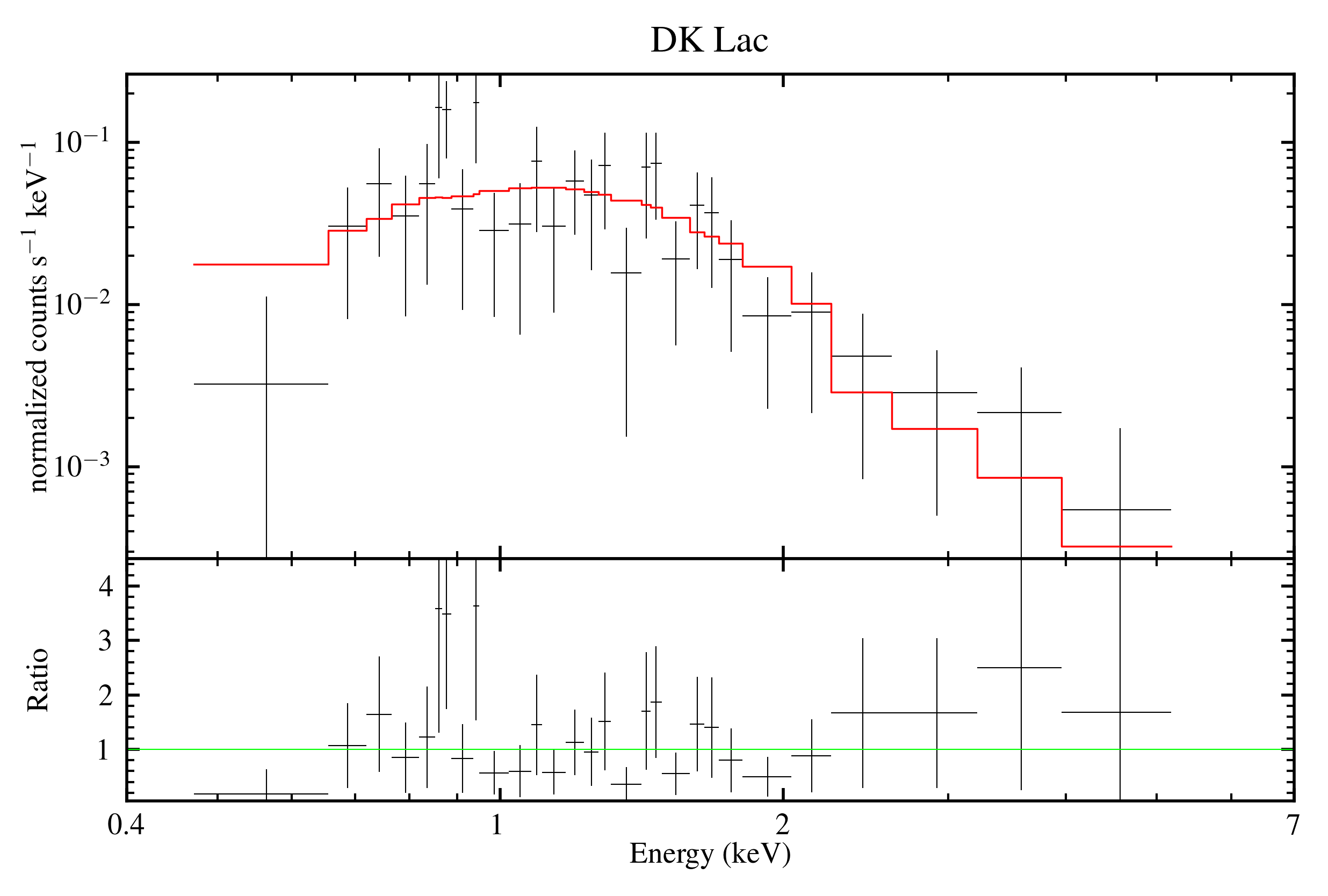}
}
\hbox
{
\includegraphics[width=60mm]{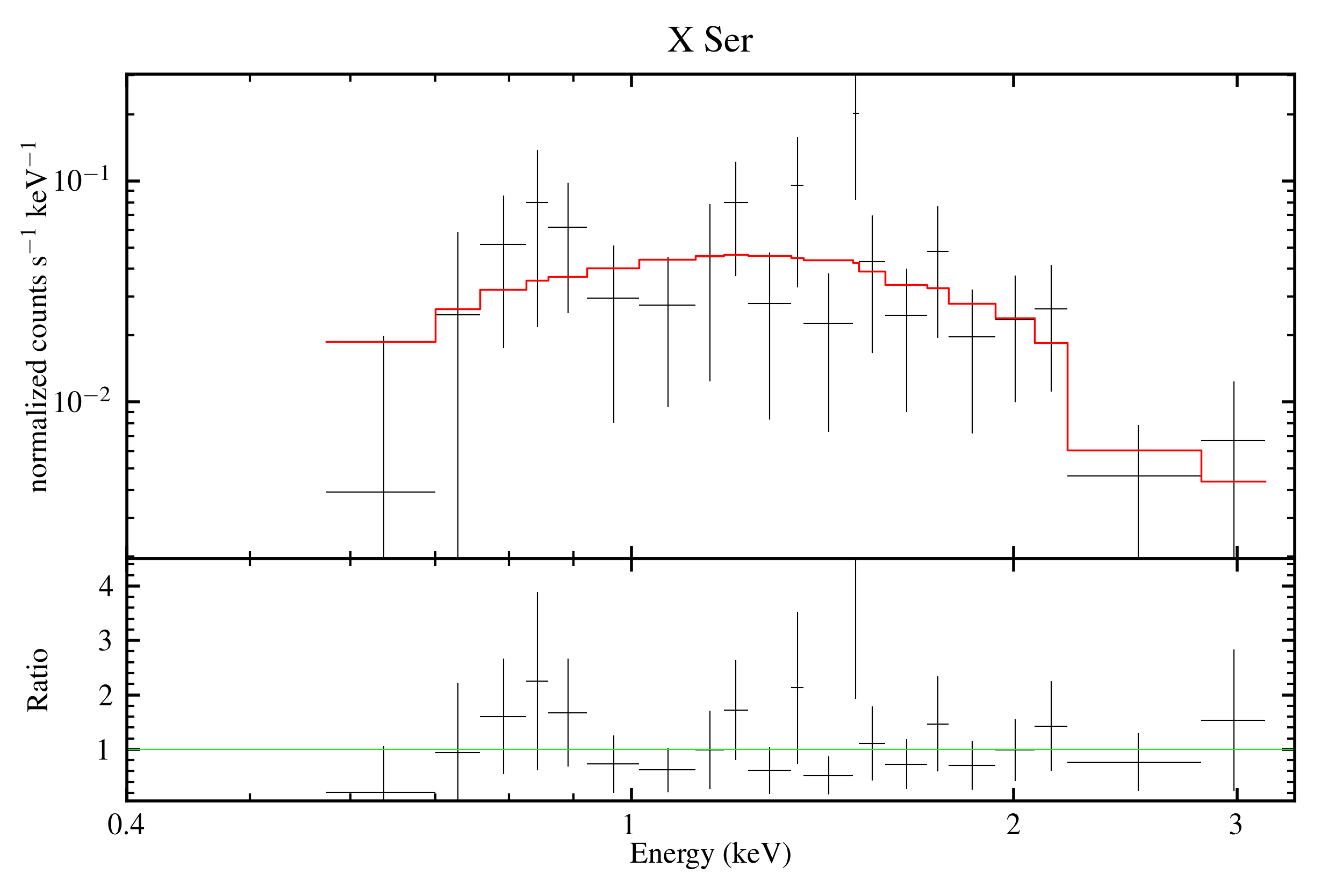}
\includegraphics[width=60mm]{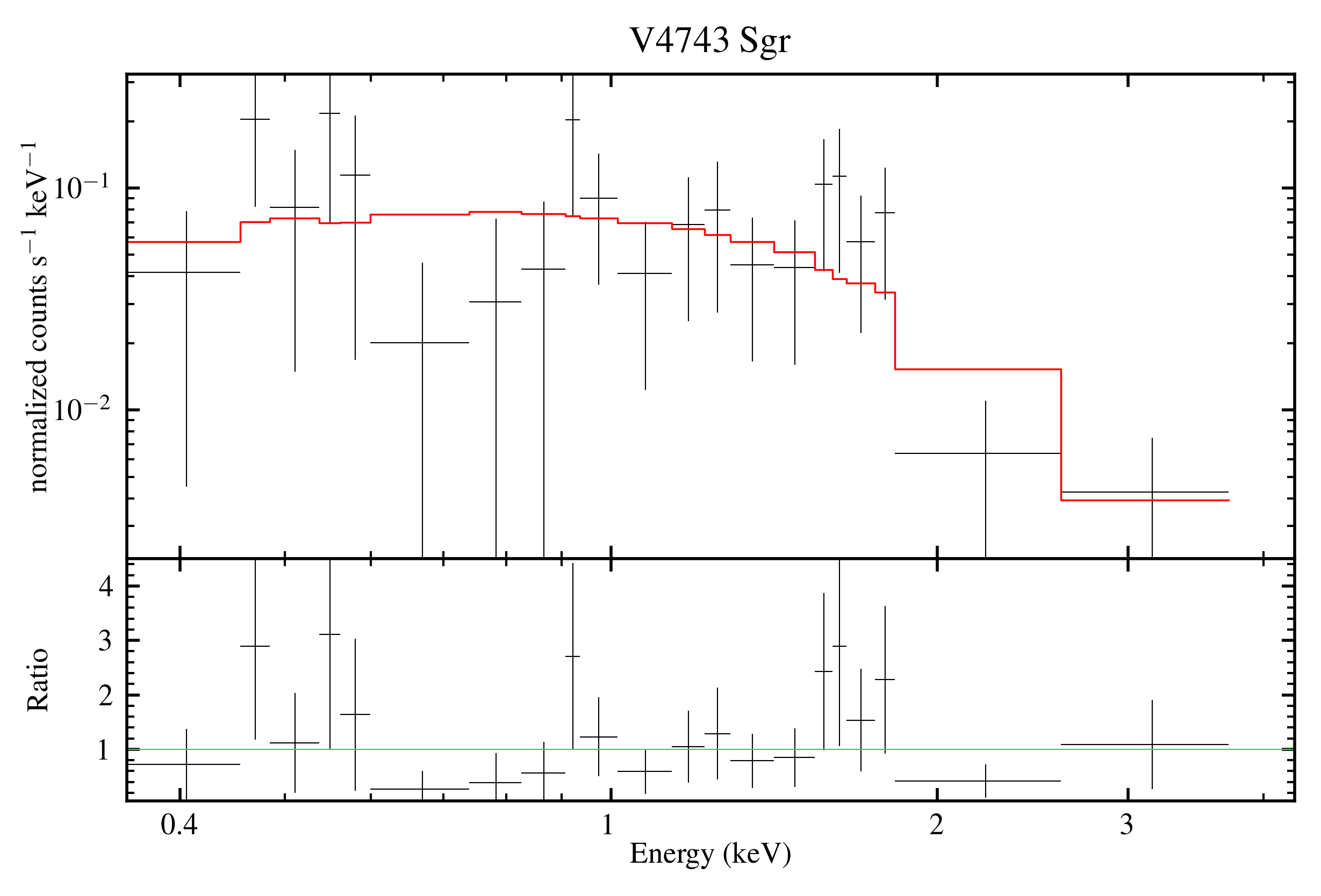}
\includegraphics[width=60mm]{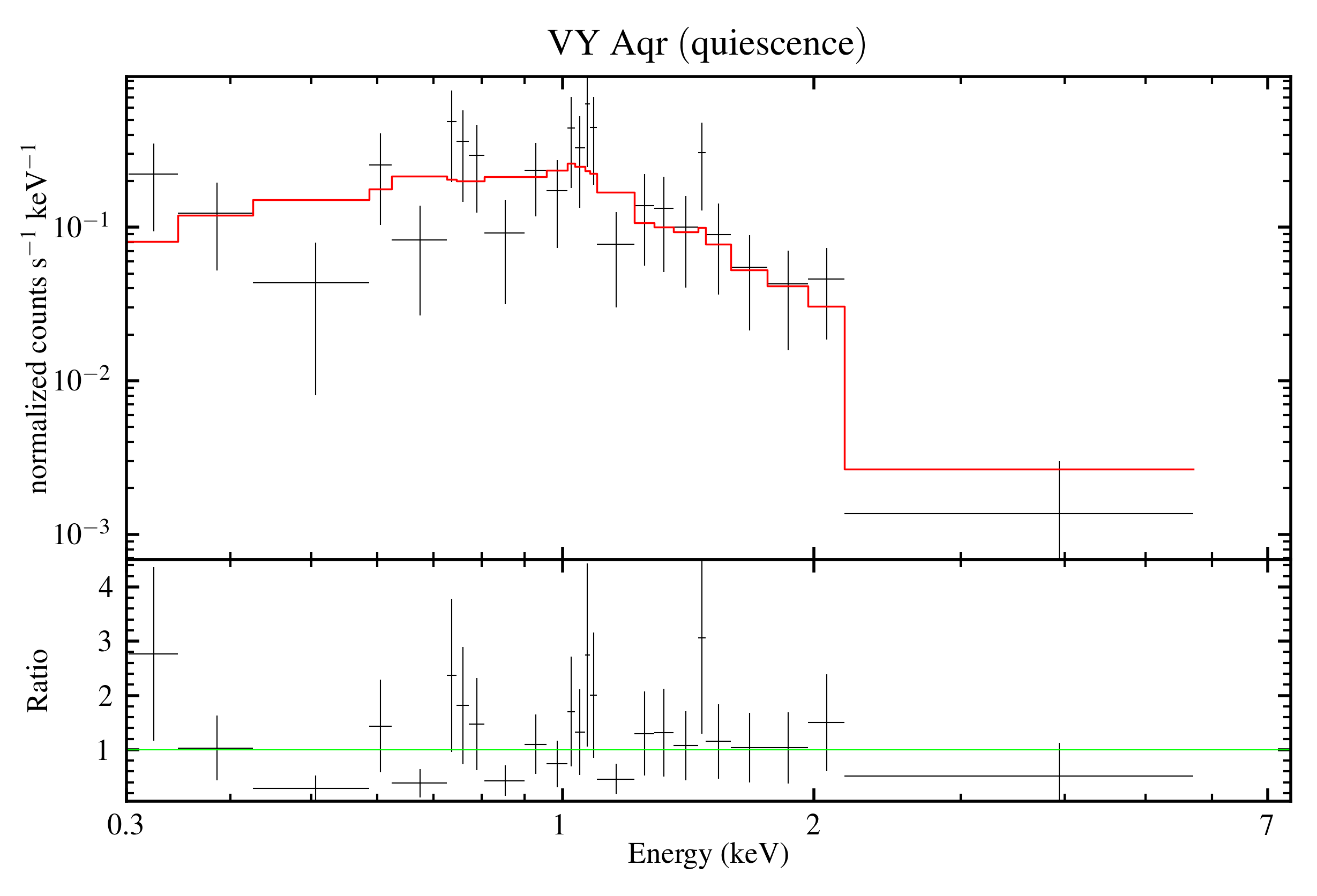}
\includegraphics[width=60mm]{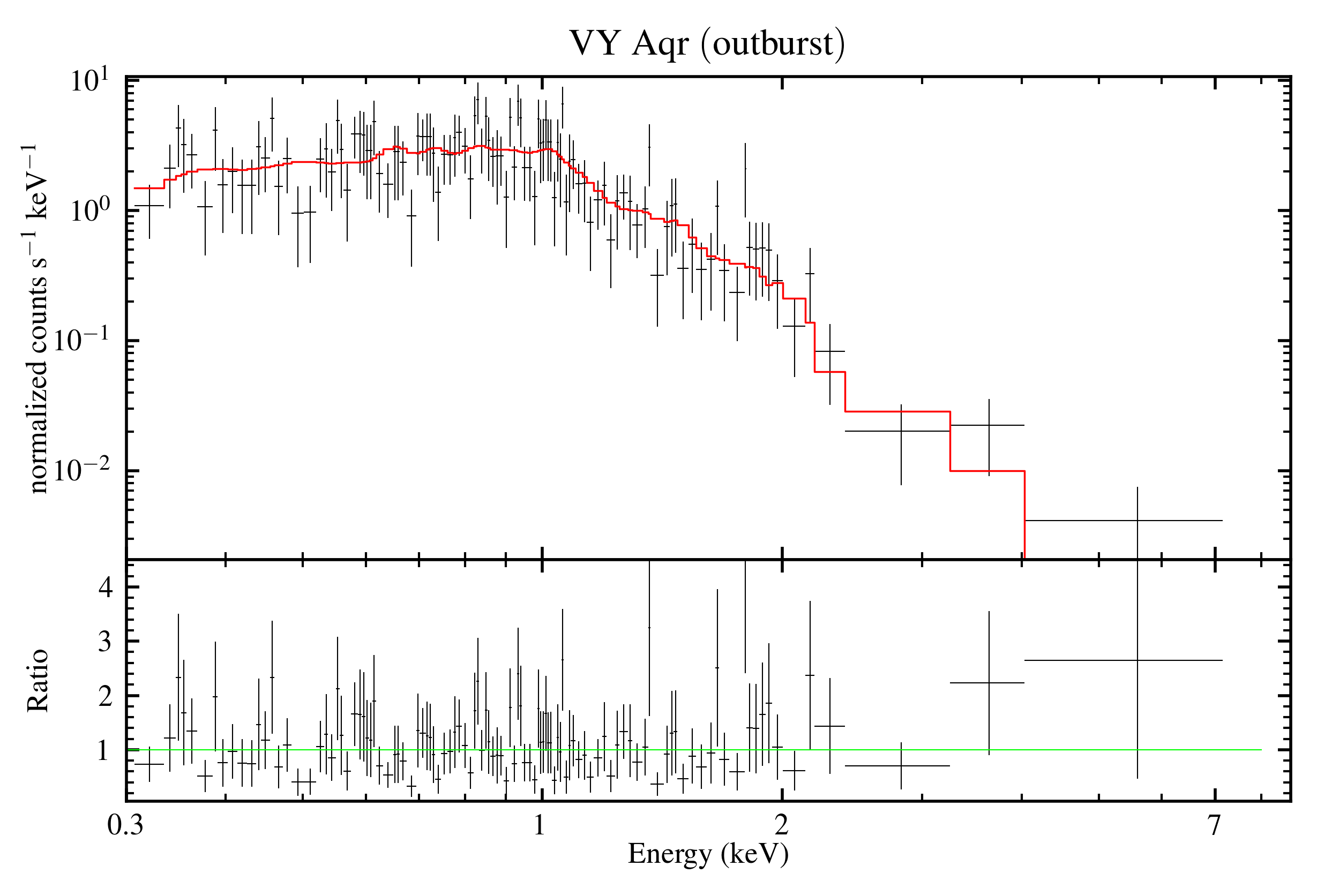}
}

\caption{\label{fig:brights_spectra} X-ray spectra of bright sources, where SRG/eROSITA detected more than 50 counts. The best-fit model selected by the Akaike Information Criterion (AIC) from Table \ref{tab:spectra_list}, presented in red. The smaller bottom panel in each spectrum shows the ratio of the observed counts to the model spectra.}

\end{center}
\end{figure*}
\end{landscape}

(2) We analyzed in detail the spectra of the bright sources from which SRG/eROSITA recorded more than 50 counts (11 sources, see Fig. \ref{fig:brights_spectra} and Table \ref{tab:spectra_list}). The bright sources can be divided by the properties of their spectra in the 0.3–7 keV energy band into two groups. For most (7 of the 11) sources, the spectra are best fitted by the isobaric cooling flow model with parameters typical for cataclysmic variables. For four sources, GK Per, V392 Per, X Ser, and V4743 Sgr, the isobaric cooling flow model is inapplicable. These sources have anomalously hard spectra with a photon index $\Gamma\sim 1$, whereas $\Gamma\sim$ 2--3 for the seven sources mentioned above. Among these four sources, GK Per and V4743 Sgr, are an intermediate polar and a candidate for intermediate polars. We hypothesized that the two remaining sources could also be intermediate polars, but the available data are not enough to test this hypothesis.

(3) The averaged spectrum of the faint sources (having less than 50 counts in their spectra) in the 0.3–7 keV energy band is anomalously hard and is well fitted by a single component power-law model with a photon index
$\rm \Gamma=0.96^{+0.18}_{-0.17}$ (see Fig. \ref{fig:comb_spectra}), resembling the spectra of intermediate
polars. The interpretation of this result requires further observations, in particular, during future SRG/eROSITA sky surveys.

(4) The historical CNe represent a sample of bona fide accreting WD with unstable thermonuclear hydrogen burning on their surface. Their X-ray luminosities in quiescence are a good proxy for the accretion
rate in a binary system. We have used this fact for the first time to construct the accretion rate distribution of WDs with non-steady hydrogen burning on their surface and compared it with the accretion rate distribution for steady supersoft sources, where hydrogen burning is steady. The derived distributions (see \ref{fig:Mdot_panels}) occupy nonintersecting regions in accretion rate and quantitatively agree with the predictions of the theory of thermonuclear hydrogen burning on the WD surface.

(5) In our sample of historical CNe, we found four sources with supersoft spectra from which there is no emission above $\rm \sim 1$ keV. For two of them, V5558 Sgr and V3666 Oph, the parameters of the black body fit, in particular, the sizes of the emitting surface, suggest that we are dealing with the post-nova supersoft X-ray emission observed $\rm \approx 13$  and $\rm \approx 1.6$ years after the CN outburst. One source, V1500 Cyg, is a known polar, and the size of the emitting surface in its case is much smaller than the WD size and is close to the expected size of the circumpolar region of a magnetized WD. The fourth source, V5856 Sgr, has the same supersoft emission parameters, suggesting that it might also be a polar.

(6) A supersoft component with similar parameters is also detected in the spectrum of V2491 Cyg, along with a harder spectral component. As has already been proposed in earlier papers \citep[see, e.g.,][]{2015ApJ...807...61Z}, the parameters of the supersoft component also suggest that this source might be an intermediate polar.

(7) During the second sky survey, in November 2020, the SRG/eROSITA telescope observed a dwarf nova outburst in VY Aqr, which enters into our CN sample. By chance coincidence, the X-ray observations occurred at the peak of the optical light curve. We analyzed in detail the X-ray spectra of this source in its outburst and in quiescence.

(8) On the whole, our X-ray spectroscopy suggests that systems with magnetized WDs may account for some fraction in our sample of historical CNe detected in X-rays. This suggestion will be checked during further SRG/eROSITA sky surveys. More detailed optical studies of these sources are also of great importance.

\acknowledgements

This work is based on observations with eROSITA telescope onboard SRG observatory. The SRG observatory was built by Roskosmos in the interests of the Russian Academy of Sciences represented by its Space Research Institute (IKI) in the framework of the Russian Federal Space Program, with the participation of the Deutsches Zentrum für Luft- und Raumfahrt (DLR). The SRG/eROSITA X-ray telescope was built by a consortium of German Institutes led by MPE, and supported by DLR.  The SRG spacecraft was designed, built, launched and is operated by the Lavochkin Association and its subcontractors. The science data are downlinked via the Deep Space Network Antennae in Bear Lakes, Ussurijsk, and Baykonur, funded by Roskosmos. The eROSITA data used in this work were processed using the eSASS software system developed by the German eROSITA consortium and proprietary data reduction and analysis software developed by the Russian eROSITA Consortium. The authors acknowledge partial support of this work by the RSF grant 19-12-00369.

\bibliographystyle{astl}
\bibliography{nova_paper}

  \onecolumn

  \fontsize{6}{10}\selectfont
       \onecolumn
       \begin{landscape}
       \renewcommand{\arraystretch}{1.0}
       \renewcommand{\tabcolsep}{2.87pt}
       \begin{longtable}{l|rrrrrrrrrr|rr}
       \caption{\label{tab:spectra_list} Best fits to the X-ray spectra of the bright CN counterparts from which the SRG/eROSITA telescope detected more than 50 counts.} \\
       \hline
       {\it Models}                     & V603 Aql                  & GK Per                    & WZ Sge                               & V2487 Oph                         & V392 Per                             &V2491 Cyg                            & WX Cet                     & DK Lac                                      & X Ser                       & V4743 Sgr                  & VY Aqr                     & VY Aqr                      \\
                                        &                           &                           &                                      &                                   &                                      &                                     &                            &                                             &                             &                            & (quiescence)                 & (outburst)                  \\
       \hline
       $\it tbabs \times power-law$     &                           &                           &                                      &                                   &                                      &                                     &                            &                                             &                             &                            &                            &                              \\
    \hline
     $ N_H[\times 10^{21}$ cm$\rm^{-2}$] &$\rm 1.52\pm{0.28}$        &$\rm 2.51_{-0.52}^{+0.57}$ &$\rm 0.41_{-0.31}^{+0.34}$            &$\rm 1.57_{-0.52}^{+0.62}$         &$\rm 1.37\pm{0.71}$                   &$\rm 0.91_{-0.73}^{+0.93}$           &$\rm 1.01_{-0.75}^{+0.99}$  &$\rm 4.30_{-2.78}^{+3.98}$                  &$\rm \le5.04$                 &$\rm \le2.34$               &$\rm \le 1.1$               & $\rm 0.96\pm{0.30}$          \\
       $\Gamma$                         &$\rm 2.27\pm{0.17}$        &$\rm 1.01\pm{0.19}$        &$\rm 1.80_{-0.26}^{+0.24}$            &$\rm 2.26_{-0.34}^{+0.42}$         &$\rm 0.96\pm{0.26}$                   &$\rm 3.20_{-0.76}^{+0.99}$           &$\rm 2.45_{-0.59}^{+0.71}$  &$\rm 2.70_{-1.32}^{+2.01}$                  &$\rm 1.15_{-1.25}^{+1.82}$    &$\rm 1.20_{-0.48}^{+1.36}$  &$\rm 1.59_{-0.38}^{+0.53}$  & $\rm 2.61\pm{0.24}$          \\
       $ C-stat/d.o.f.$              &317.8/248                  &252.0/199                  &126.2/138                             &143.6/119                          &84.1/72                               &89.9/50                              &42.2/45                     &23.8/26                                     &12.8/18                       &18.7/18                     &31.6/22                     & 141.7/108                    \\
       $ \alpha$(goodness)           &$\rm \le10^{-5}$           &$\rm 10^{-2}$              &$\rm 0.57$                            &$\rm \le10^{-5}$                   &$\rm 0.32$                            &$\rm \le10^{-5}$                     &$\rm 0.54$                  &$\rm 0.35$                                  &$\rm 0.73$                    &$\rm 0.48$                  &$\rm 0.11$                  & $\rm \le10^{-5}$              \\
       $ AIC$                        &323.8                      &258.0                      &132.2                                 &149.6                              &90.2                                  &95.9                                 &48.2                        &29.8                                        &18.8                          &24.7                        &37.6                        & 147.7                        \\
       $ log(L_{X}^{pw}$[erg/s]$\rm)^a$   &$\rm 32.37_{-0.04}^{+0.06}$&$\rm 32.50\pm{0.04}$       &$\rm 30.00\pm{0.03}$                  &$\rm 35.54_{-0.08}^{+0.13}$        &$\rm 33.81_{-0.09}^{+0.10}$           &$\rm 34.46_{-0.34}^{+0.50}$          &$\rm 30.25_{-0.15}^{+0.28}$ &$\rm 32.73_{-0.41}^{+0.84}$                 &$\rm 33.09_{-0.15}^{+0.30}$   &$\rm 33.13\pm{0.16}$        &$\rm 30.08_{-0.09}^{+0.14}$ & $\rm 31.07_{-0.12}^{+0.16}$  \\
       \hline
       $\it tbabs \times meka $         &                           &                           &                                      &                                   &                                      &                                     &                            &                                            &                              &                            &                            &                               \\
      \hline
    $ N_H\ [\times 10^{21}$ cm$\rm^{-2}$] &$\rm 0.89_{-0.15}^{+0.09}$ & $\rm 3.30\pm{0.21}$       &$\rm \le 0.02$                        &$\rm 0.71\pm{0.20}$                &$\rm 2.35_{-0.36}^{+0.38}$            &$\rm \le0.40$                        &$\rm \le1.37$               &$\rm 2.96_{-1.34}^{+1.56}$                  &5.50                          &$\rm \le0.10$               &$\rm \le 0.87$              & $\rm 0.45\pm{0.15}$            \\
     $ kT$ [keV]                     &$\rm 2.02_{-0.21}^{+0.30}$ & $\rm 10^*$                &$\rm 4.62_{-1.58}^{+1.37}$            &$\rm 3.05_{-0.64}^{+1.53}$         & $\rm 10^*$                           &$\rm 0.97_{-0.22}^{+0.28}$           &$\rm 1.37_{-0.30}^{+0.61}$  &$\rm 1.67_{-0.88}^{+5.53}$                  &$\rm 1.52_{-0.50}^{+1.10}$    &$\rm 10^*$                  &$\rm 1.36_{-0.29}^{+1.41}$  & $\rm 0.80_{-0.06}^{+0.09}$     \\
     $ Z[\ Z_{\odot}]$               & $\rm \le0.02$             & $\rm \le2$                &$\rm 0.40_{-0.37}^{+0.61}$            &$\rm \le0.1$                       &$\rm \le3.2$                          &$\rm \le5\times10^{-3}$              &$\rm 0.16_{-0.11}^{+0.27}$  &$\rm \le 0.2$                               &$\rm \le0.14$                 &$\rm \le 2$                 &$\rm 0.38_{-0.24}^{+0.87}$  & $\rm 0.06_{-0.02}^{+0.03}$     \\
       $ C-stat/d.o.f.$              &314.2/247                  &264.4/199                  &125.8/137                             &154.6/118                          &90.0/72                               &106.5/49                             &36.8/44                     &24.4/25                                     &14.7/17                       &18.8/18                     &26.6/21                     & 126.4/107                      \\
       $ \alpha$(goodness)           &$\rm \le10^{-5}$           &$\rm 9\times 10^{-3}$      &$\rm 0.58$                            &$\rm 10^{-2}$                      &$\rm 0.12$                            &$\rm \le10^{-5}$                     &$\rm 0.68$                  &$\rm 0.67$                                  &$\rm 0.56$                    &$\rm 0.43$                  &$\rm 0.08$                  & $\rm 0.11$                     \\
       $ AIC$                        &322.2                      &272.4                      &133.8                                 &162.6                              &95.0                                  & 114.5                               &44.8                        &32.4                                        &22.7                          & 26.8                       &34.6                        & 134.4                           \\
    $ log(L_{X}^{meka}$[erg/s]$\rm)^a$    &$\rm 32.29_{-0.02}^{+0.01}$&$\rm 32.52\pm{0.03}$       &$\rm 30.00\pm{0.07}$                  &$\rm 35.49\pm{0.04}$               &$\rm 33.80\pm{0.06}$                  &$\rm 34.30_{-0.04}^{+0.05}$          &$\rm 30.12\pm{0.08}$        &$\rm 32.54_{-0.10}^{+0.22}$                 &$\rm 33.10\pm{0.08}$          &$\rm 33.13\pm{0.10}$        &$\rm 29.92\pm{0.10}$        & $\rm 30.96\pm{0.04}$            \\
       \hline
       $\it tbabs \times mkcflow $      &                           &                           &                                      &                                   &                                      &                                     &                            &                                            &                              &                            &                            &                                   \\
       \hline
     $ N_H[\times 10^{21}$ cm$\rm^{-2}$] &$\rm 0.73\pm{0.13}$        &                           &$\rm \le 0.02$                        &$\rm \le 0.13$                     &                                      &$\rm \le 0.29$                       &$\rm \le0.27$               &$\rm 2.93_{-0.86}^{+1.23}$                  &                              &                            &$\rm \le 0.37$              & $\rm \le 0.2$                     \\
     $ kT_{min}$ [keV]                &$\rm \le 0.19$             &                           &$\rm \le 1.06$                        &$\rm \le 0.92$                     &                                      &$\rm \le 0.25$                       &$\rm 8^*\times 10^{-3}$     &$\rm \le 0.91$                              &                              &                            &$\rm 0.95_{-0.75}^{+1.29}$  & $\rm \le 0.31$                         \\
       $ kT_{max}$ [keV]              &$\rm 6.97_{-1.30}^{+1.53}$ &                           &$\rm 17.20_{-6.94}^{+20.33}$          &$\rm 24.72_{-9.91}^{+22.63}$       &                                      &$\rm 1.98_{-0.70}^{+0.93}$           &$\rm 6.37_{-2.51}^{+2.71}$  &$\rm 5.15_{-2.69}^{+18.69}$                 &                              &                            &$\rm 5.14_{-3.29}^{+8.35}$  & $\rm 2.91_{-0.40}^{+0.45}$                \\
       $ Z[\ Z_{\odot}]$             &$\rm 0.19_{-0.08}^{+0.10}$ &                           &$\rm 0.32_{-0.19}^{+0.34}$            &$\rm \le0.21$                      &                                      & $\rm \le 3\times10^{-3}$            &$\rm 0.55_{-0.41}^{+0.59}$  &$\rm \le 0.35$                              &                              &                            &$\rm 1^*$                   & $\rm 0.24_{-0.08}^{+0.09}$        \\
    $ \dot{M}_{acc}
    [\times 10^{-12}\ M_\odot$/yr]     &$\rm 53.3_{-9.8}^{+15.2}$  &                           &$\rm 11.0_{-3.9}^{+5.0}\times10^{-2}$ &$\rm 3.1_{-1.0}^{+1.1}\times10^{4}$&                                      &$\rm 2.3_{-0.8}^{+2.3}\times10^{4}$  &$\rm 0.4_{-0.1}^{+0.2}$     &$\rm 135.1_{-135.1}^{+983.1}$               &                              &                            &$\rm 0.5_{-0.3}^{+1.6}$      & $\rm 6.6_{-0.9}^{+2.3}$         \\
        $ C-stat/d.o.f.$             &287.9/246                  &                           &123.5/137                             &113.5/116                         &                                      &101.6/49                             &37.3/44                     &23.3/24                                     &                              &                            &24.6/21                     & 109.6/107                          \\
      $ \alpha$(goodness)            &$\rm 0.11$                 &                           &$\rm 0.54$                            &$\rm 0.21$                         &                                      &$\rm \le10^{-5}$                     & $\rm 0.66$                 & $\rm 0.55$                                 &                              &                            &$\rm 0.19$                  & $\rm 0.45$                          \\
        $ AIC$                       &297.9                      &                           &131.5                                 &123.52                             &                                      &101.6                                & 45.3                       & 33.3                                       &                              &                            &31.7                        & 119.6                               \\
  $ log(L_{X}^{mkcflow}$[erg/s]$\rm)^a$    &$\rm 32.23\pm{0.02}$       &                           &$\rm 29.86\pm{0.05}$                  &$\rm 35.45\pm{0.06}$               &                                      &$\rm 34.20\pm{0.04}$                 &$\rm 30.09\pm{0.07}$        &$\rm 32.49_{-0.11}^{+0.33}$                 &                              &                            &$\rm 29.97_{-0.11}^{+0.13}$ & $\rm 30.91_{-0.03}^{+0.04}$         \\

      \hline
     $\it tbabs \times (bbody+meka) $   &                           &                           &                                      &                                   &                                       &                                    &                            &                                            &                              &                            &                            &                 \\
      \hline
    $ N_H[\times 10^{21}$ cm$\rm^{-2}$]  &                           &                           &                                      &$\rm 4.25_{-1.41}^{+1.53}$         &                                       &$\rm 3.61_{-1.55}^{+1.24}$          &                            &                                            &                              &                            &                            &                 \\
     $ kT_{BB}$ [eV]                 &                           &                           &                                      &$\rm 96_{-14}^{+21}$               &                                       &$\rm 65_{-9}^{+11}$                 &                            &                                            &                              &                            &                            &                 \\
     $ kT_{meka}$ [keV]               &                           &                           &                                      &$\rm 6.7_{-2.7}^{+48.8}$           &                                       & $\rm 10^*$                         &                            &                                            &                              &                            &                            &                 \\
     $ Z[\ Z_{\odot}]$               &                           &                           &                                      & $\rm 1^*$                         &                                       & $\rm 1^*$                          &                            &                                            &                              &                            &                            &                 \\
       $ C-stat/d.o.f.$              &                           &                           &                                      &116.7/117                          &                                       &44.9/49                             &                            &                                            &                              &                            &                            &                 \\
      $ \alpha$(goodness)            &                           &                           &                                      &$\rm 0.82$                         &                                       &$\rm 0.67$                          &                            &                                            &                              &                            &                            &                 \\
       $ AIC$                        &                           &                           &                                      &127.0                              &                                       &52.9                                &                            &                                            &                              &                            &                            &                 \\
       $ log(L_{bol}^{BB}$[erg/s]$\rm)^a$  &                           &                           &                                      &$\rm 36.31\pm{0.61}$               &                                       &$\rm 36.05_{-0.73}^{+0.81}$         &                            &                                            &                              &                            &                            &                 \\
     $ log(L_{X}^{meka}$[erg/s]$\rm)^a$    &                           &                           &                                      &$\rm 35.54\pm{0.12}$               &                                       &$\rm 34.41\pm{0.12}$                &                            &                                            &                              &                            &                            &                 \\
     $ R_{BB}$ [km]                 &                           &                           &                                      &$\rm 439_{-289}^{+618}$            &                                       &$\rm 700_{-458}^{+1552}$            &                            &                                            &                              &                            &                            &                 \\
       \hline
       \end{longtable}
 Notes: * -- The parameter was fixed at a given value. (a) Absorption corrected X-ray luminosity in the 0.3-8 keV energy band. Distances from Table \ref{tab:nova_list} were used to calculate luminosities. Error bars show 1$\sigma$ confidence limits.
       \end{landscape}

\fontsize{6.5}{11}\selectfont

\onecolumn

\begin{landscape}
\renewcommand{\arraystretch}{1.2}
\renewcommand{\tabcolsep}{3.2pt}
\begin{longtable}{lcrrccccccrrrr}
\caption{\label{tab:nova_list} List of X-ray sources from the SRG/eROSITA catalog having identifications with CNe in our Galaxy.} \\
\hline
\textnumero & Date & Nova & RA & DEC &$ R_{\rm err}(98\%)$ & SRGe &\arcsec & $ N_{\rm H,Gal}$ & E(B-V) & $\rm d$  & $ M_{\rm WD}$& $ L_X$ & $ L^0_X$  \\
& &  &(\degr)  &(\degr)  &(\arcsec)  &  &  &($\rm 10^{21}$ cm$\rm^{-2}$)  & &(kpc) &($\rm M_\odot$)  & (erg/s) &(erg/s)  \\
(1) & (2) & (3) & (4)  &(5) & (6) & (7) & (8) & (9) & (10) & (11) & (12) & (13) & (14)\\
\hline
      1 &   12/06/2021 (1) &    V1674 Her &   284.37953 &     16.89338 &             6.87 &   J185731.1+165336 &         2.29 &                  2.99 &                 0.72 (1) &                      --  &                     --  &  $\rm (2.2\pm 0.6)\times 10^{31}$ &  $\rm (4.8\pm 1.2)\times 10^{31}$ \\
     2 &   18/03/2021 (2) &    V1405 Cas &   351.20222 &     61.18732 &             9.88 &   J232448.5+611114 &         4.42 &                  7.07 &   $\rm 0.56\pm0.02\ (3)$ &   $\rm 1.73\pm0.01\ (G)$ &                     --  &  $\rm (6.1\pm 2.4)\times 10^{30}$ &  $\rm (1.2\pm 0.5)\times 10^{31}$ \\
     3 &   29/10/2019 (4) &     V659 Sct &   279.99787 &    -10.42678 &             6.04 &   J183959.5-102536 &         6.26 &                  5.37 &                  0.9 (5) &   $\rm 1.97\pm0.66\ (G)$ &                     --  &  $\rm (7.8\pm 1.6)\times 10^{31}$ &  $\rm (1.9\pm 0.4)\times 10^{32}$ \\
     4 &   27/08/2019 (6) &    V3890 Sgr &   277.68084 &    -24.01880 &            13.80 &   J183043.4-240108 &        10.92 &                  1.64 &                 0.57 (7) &                  4.5 (7) &                 1.3 (a) &  $\rm (2.0\pm 0.6)\times 10^{32}$ &  $\rm (3.9\pm 1.2)\times 10^{32}$ \\
     5 &   08/08/2018 (8) &    V3666 Oph &   265.60063 &    -20.88580 &             9.67 &   J174224.2-205309 &         1.14 &                  2.88 &                 0.98 (9) &                      --  &                     --  &  $\rm (1.5\pm 0.5)\times 10^{31}$ &  $\rm (3.8\pm 1.2)\times 10^{31}$ \\
     6 &  29/04/2018 (10) &     V392 Per &    70.84108 &     47.35684 &             3.37 &   J044321.9+472125 &         1.88 &                  5.82 &                0.72 (11) &                3.88 (11) &                1.21 (b) &  $\rm (1.3\pm 0.1)\times 10^{33}$ &  $\rm (2.8\pm 0.2)\times 10^{33}$ \\
     7 &  19/06/2017 (12) &     V612 Sct &   277.94124 &    -14.31627 &             8.41 &   J183145.9-141859 &         2.38 &                  4.37 &                0.68 (13) &   $\rm 3.99\pm1.0\ (14)$ &                     --  &  $\rm (1.9\pm 0.5)\times 10^{32}$ &  $\rm (3.9\pm 1.1)\times 10^{32}$ \\
     8 &  25/10/2016 (15) &    V5856 Sgr &   275.22035 &    -28.37153 &             7.19 &   J182052.9-282218 &        11.59 &                  1.43 &                0.34 (16) &                 4.2 (16) &                     --  &  $\rm (2.9\pm 0.6)\times 10^{32}$ &  $\rm (4.7\pm 1.0)\times 10^{32}$ \\
     9 &  14/08/2013 (17) &     V339 Del &   305.87688 &     20.76760 &             6.20 &   J202330.5+204603 &         1.98 &                  1.30 &                0.18 (18) &    $\rm 4.5\pm0.8\ (18)$ &                     --  &  $\rm (2.8\pm 0.6)\times 10^{32}$ &  $\rm (3.8\pm 0.8)\times 10^{32}$ \\
    10 &  08/05/2012 (19) &    V5590 Sgr &   272.76528 &    -27.29020 &             8.69 &   J181103.7-271725 &         5.57 &                  2.47 &                      --  &                      --  &                     --  &  $\rm (3.0\pm 0.8)\times 10^{31}$ &  $\rm (4.9\pm 1.3)\times 10^{31}$ \\
    11 &  07/11/2011 (20) &     V965 Per &    47.81798 &     37.08439 &            12.56 &   J031116.3+370504 &         2.37 &                  1.09 &                      --  &                      --  &                     --  &  $\rm (5.2\pm 2.4)\times 10^{30}$ &  $\rm (6.8\pm 3.2)\times 10^{30}$ \\
    12 &  10/04/2008 (21) &    V2491 Cyg &   295.75797 &     32.31973 &             3.52 &   J194301.9+321911 &         1.08 &                  3.96 &                0.45 (22) &                15.9 (22) &                1.35 (c) &  $\rm (1.3\pm 0.1)\times 10^{34}$ &  $\rm (2.3\pm 0.2)\times 10^{34}$ \\
    13 &  14/04/2007 (23) &    V5558 Sgr &   272.57574 &    -18.78132 &             6.79 &   J181018.2-184653 &         1.03 &                 14.81 &                 0.8 (24) &    $\rm 1.3\pm0.3\ (25)$ &   $\rm 0.6\pm0.03\ (d)$ &  $\rm (2.8\pm 0.6)\times 10^{31}$ &  $\rm (6.2\pm 1.4)\times 10^{31}$ \\
    14 &  15/03/2007 (26) &    V2467 Cyg &   307.05124 &     41.80980 &             7.76 &   J202812.3+414835 &         1.85 &                 12.52 &                 1.5 (27) &    $\rm 3.1\pm0.5\ (28)$ &  $\rm 1.04\pm0.07\ (e)$ &  $\rm (1.3\pm 0.3)\times 10^{32}$ &  $\rm (4.1\pm 0.9)\times 10^{32}$ \\
    15 &  12/02/2006 (29) &       RS Oph &   267.55374 &     -6.70904 &             7.08 &   J175012.9-064233 &         3.92 &                  2.08 &                0.73 (30) &    $\rm 1.6\pm0.3\ (31)$ &                1.35 (c) &  $\rm (2.8\pm 0.7)\times 10^{31}$ &  $\rm (6.1\pm 1.4)\times 10^{31}$ \\
    16 &  20/09/2002 (32) &    V4743 Sgr &   285.28921 &    -22.00149 &             4.93 &   J190109.4-220005 &         1.71 &                  1.02 &                0.25 (33) &    $\rm 3.9\pm0.3\ (33)$ &  $\rm 1.15\pm0.06\ (e)$ &  $\rm (4.8\pm 0.7)\times 10^{32}$ &  $\rm (6.9\pm 1.0)\times 10^{32}$ \\
    17 &  15/06/1998 (34) &    V2487 Oph &   262.99910 &    -19.23298 &             2.89 &   J173159.8-191359 &         1.63 &                  2.03 &                0.38 (35) &   $\rm 27.5\pm3.0\ (35)$ &  $\rm 1.35\pm0.01\ (f)$ &  $\rm (1.4\pm 0.1)\times 10^{35}$ &  $\rm (2.4\pm 0.1)\times 10^{35}$ \\
    18 &  19/02/1992 (36) &    V1974 Cyg &   307.62953 &     52.62978 &             7.79 &   J203031.1+523747 &         2.45 &                  2.96 &                0.36 (30) &    $\rm 1.8\pm0.1\ (30)$ &                     --  &  $\rm (1.3\pm 0.4)\times 10^{31}$ &  $\rm (2.1\pm 0.7)\times 10^{31}$ \\
    19 &  21/03/1988 (37) &       PQ And &    37.37428 &     40.04330 &             6.15 &   J022929.8+400236 &         4.18 &                  0.52 &                      --  &   $\rm 0.27\pm0.02\ (G)$ &                     --  &  $\rm (8.0\pm 1.5)\times 10^{29}$ &  $\rm (9.3\pm 1.8)\times 10^{29}$ \\
    20 &  21/10/1976 (38) &       NQ Vul &   292.31150 &     20.46575 &            10.46 &   J192914.8+202757 &         4.23 &                 10.07 &   $\rm 0.92\pm0.2\ (39)$ &    $\rm 1.6\pm0.8\ (40)$ &                     --  &  $\rm (1.5\pm 0.5)\times 10^{31}$ &  $\rm (3.7\pm 1.3)\times 10^{31}$ \\
    21 &  29/08/1975 (41) &    V1500 Cyg &   317.90211 &     48.15137 &             6.06 &   J211136.5+480905 &         4.14 &                  8.76 &                0.45 (22) &                 1.5 (22) &                 1.2 (c) &  $\rm (2.9\pm 0.6)\times 10^{31}$ &  $\rm (5.1\pm 1.1)\times 10^{31}$ \\
    22 &  10/07/1971 (42) &       IV Cep &   331.15760 &     53.50736 &            11.10 &   J220437.8+533027 &         8.71 &                  5.42 &                0.65 (22) &                 3.1 (22) &                0.98 (c) &  $\rm (2.3\pm 0.9)\times 10^{31}$ &  $\rm (4.6\pm 1.9)\times 10^{31}$ \\
    23 &   15/04/1968 (C) &       LV Vul &   297.00245 &     27.17116 &             8.24 &   J194800.6+271016 &         2.69 &                  9.62 &                 0.6 (22) &                 1.0 (22) &                0.98 (c) &  $\rm (1.2\pm 0.3)\times 10^{31}$ &  $\rm (2.3\pm 0.5)\times 10^{31}$ \\
    24 &  08/07/1967 (43) &       HR Del &   310.58375 &     19.15977 &             7.99 &   J204220.1+190935 &         3.59 &                  0.71 &  $\rm 0.17\pm0.02\ (44)$ &   $\rm 0.96\pm0.03\ (G)$ &  $\rm 0.67\pm0.08\ (g)$ &  $\rm (5.6\pm 1.8)\times 10^{30}$ &  $\rm (7.4\pm 2.4)\times 10^{30}$ \\
    25 &   04/11/1964 (C) &       QZ Aur &    82.14550 &     33.30623 &             9.89 &   J052834.9+331822 &         8.20 &                  4.84 &                0.55 (45) &   $\rm 2.87\pm0.89\ (G)$ &                     --  &  $\rm (2.9\pm 1.3)\times 10^{31}$ &  $\rm (5.5\pm 2.4)\times 10^{31}$ \\
    26 &   14/09/1963 (C) &       AS Psc &    22.03717 &     31.24893 &            16.82 &   J012808.9+311456 &        13.98 &                  0.50 &                      --  &                      --  &                     --  &  $\rm (5.0\pm 2.1)\times 10^{30}$ &  $\rm (5.7\pm 2.5)\times 10^{30}$ \\
    27 &   01/09/1963 (C) &       WX Cet &    19.26751 &    -17.93925 &             3.55 &   J011704.2-175621 &         1.82 &                  0.15 &                      --  &                0.13 (46) &                     --  &  $\rm (8.6\pm 0.8)\times 10^{29}$ &  $\rm (9.0\pm 0.8)\times 10^{29}$ \\
    28 &   06/02/1963 (C) &     V533 Her &   273.58518 &     41.85650 &             6.55 &   J181420.4+415123 &         4.95 &                  0.35 &  $\rm 0.03\pm0.02\ (44)$ &                1.28 (22) &                1.03 (c) &  $\rm (1.8\pm 0.3)\times 10^{31}$ &  $\rm (1.9\pm 0.3)\times 10^{31}$ \\
    29 &   07/03/1960 (C) &     V446 Her &   284.34060 &     13.24122 &             7.75 &   J185721.7+131428 &         3.10 &                  3.92 &  $\rm 0.38\pm0.04\ (44)$ &                1.38 (22) &                0.98 (c) &  $\rm (2.3\pm 0.5)\times 10^{31}$ &  $\rm (3.8\pm 0.8)\times 10^{31}$ \\
    30 &   21/02/1952 (C) &    V1175 Sgr &   273.57256 &    -31.12224 &            11.16 &   J181417.4-310720 &        14.67 &                  1.57 &                0.29 (47) &                 5.2 (47) &                     --  &  $\rm (1.3\pm 0.5)\times 10^{32}$ &  $\rm (1.9\pm 0.8)\times 10^{32}$ \\
    31 &   14/11/1950 (C) &     V630 Cas &   357.21651 &     51.46181 &             5.93 &   J234852.0+512743 &         1.19 &                  1.48 &                      --  &   $\rm 3.47\pm0.63\ (G)$ &                     --  &  $\rm (1.1\pm 0.2)\times 10^{32}$ &  $\rm (1.5\pm 0.3)\times 10^{32}$ \\
    32 &   23/01/1950 (C) &       DK Lac &   342.44570 &     53.28933 &             4.80 &   J224947.0+531722 &         1.77 &                  2.70 &  $\rm 0.22\pm0.06\ (44)$ &   $\rm 2.48\pm0.43\ (G)$ &                     --  &  $\rm (1.0\pm 0.1)\times 10^{32}$ &  $\rm (1.4\pm 0.2)\times 10^{32}$ \\
    33 &   18/06/1936 (C) &       CP Lac &   333.92164 &     55.61772 &             5.37 &   J221541.2+553704 &         1.74 &                  5.84 &  $\rm 0.28\pm0.06\ (44)$ &   $\rm 1.16\pm0.06\ (G)$ &                     --  &  $\rm (1.4\pm 0.2)\times 10^{31}$ &  $\rm (2.1\pm 0.3)\times 10^{31}$ \\
    34 &   12/12/1934 (C) &       DQ Her &   271.87469 &     45.85761 &             7.01 &   J180729.9+455127 &         4.00 &                  0.32 &  $\rm 0.05\pm0.02\ (44)$ &  $\rm 0.39\pm0.03\ (14)$ &                     --  &  $\rm (1.3\pm 0.2)\times 10^{30}$ &  $\rm (1.4\pm 0.3)\times 10^{30}$ \\
    35 &   01/01/1929 (C) &       BC Cas &   357.82628 &     60.30350 &             6.26 &   J235118.3+601813 &         5.50 &                  6.31 &                      --  &   $\rm 2.04\pm0.29\ (G)$ &                     --  &  $\rm (3.0\pm 0.6)\times 10^{31}$ &  $\rm (7.3\pm 1.5)\times 10^{31}$ \\
    36 &   30/07/1927 (C) &       EL Aql &   284.00943 &     -3.32041 &            15.63 &   J185602.3-031913 &         9.36 &                  5.72 &                1.11 (48) &                      --  &                     --  &  $\rm (8.8\pm 4.3)\times 10^{30}$ &  $\rm (2.4\pm 1.2)\times 10^{31}$ \\
    37 &   14/06/1926 (C) &       KY Sgr &   270.33869 &    -26.41412 &             6.96 &   J180121.3-262451 &        12.69 &                  5.48 &                      --  &                 5.3 (47) &                     --  &  $\rm (5.2\pm 1.1)\times 10^{32}$ &  $\rm (1.2\pm 0.2)\times 10^{33}$ \\
    38 &   20/08/1920 (C) &     V476 Cyg &   299.60098 &     53.61866 &            10.38 &   J195824.2+533707 &         3.00 &                  1.77 &                0.19 (47) &                 1.8 (47) &                     --  &  $\rm (4.8\pm 2.3)\times 10^{30}$ &  $\rm (6.6\pm 3.1)\times 10^{30}$ \\
    39 &   11/03/1919 (C) &    V1017 Sgr &   278.01815 &    -29.38733 &             7.58 &   J183204.4-292314 &         1.50 &                  1.27 &  $\rm 0.39\pm0.03\ (14)$ &   $\rm 1.27\pm0.07\ (G)$ &                     --  &  $\rm (2.1\pm 0.5)\times 10^{31}$ &  $\rm (3.5\pm 0.9)\times 10^{31}$ \\
    40 &   08/06/1918 (C) &     V603 Aql &   282.22782 &      0.58420 &             2.67 &   J184854.7+003503 &         2.21 &                 14.39 &  $\rm 0.08\pm0.02\ (44)$ &   $\rm 0.31\pm0.01\ (G)$ &    $\rm 1.2\pm0.1\ (g)$ &  $\rm (1.1\pm 0.2)\times 10^{32}$ &  $\rm (1.3\pm 0.3)\times 10^{32}$ \\
    41 &   22/11/1913 (C) &       WZ Sge &   301.90201 &     17.70392 &             2.92 &   J200736.5+174214 &         5.64 &                  2.01 &                      --  &   $\rm 0.04\pm0.0\ (49)$ &  $\rm 0.85\pm0.04\ (h)$ &  $\rm (4.7\pm 0.2)\times 10^{29}$ &  $\rm (7.2\pm 0.3)\times 10^{29}$ \\
    42 &   30/12/1910 (C) &       DI Lac &   338.95185 &     52.71758 &             8.21 &   J223548.4+524303 &         2.68 &                  3.26 &  $\rm 0.26\pm0.04\ (44)$ &   $\rm 1.64\pm0.06\ (G)$ &  $\rm 0.68\pm0.12\ (g)$ &  $\rm (1.3\pm 0.3)\times 10^{31}$ &  $\rm (2.0\pm 0.5)\times 10^{31}$ \\
    43 &   21/03/1910 (C) &     V999 Sgr &   270.02649 &    -27.55126 &             9.73 &   J180006.4-273305 &        14.24 &                  4.62 &                0.58 (47) &                 2.6 (47) &                     --  &  $\rm (3.2\pm 1.3)\times 10^{31}$ &  $\rm (6.2\pm 2.5)\times 10^{31}$ \\
    44 &   12/08/1907 (C) &       VY Aqr &   318.03833 &     -8.82741 &             2.82 &   J211209.2-084939 &         4.44 &                  0.59 &                      --  &   $\rm 0.1\pm0.01\ (50)$ &                     --  &  $\rm (3.1\pm 0.1)\times 10^{30}$ &  $\rm (3.7\pm 0.2)\times 10^{30}$ \\
    45 &   01/05/1903 (C) &        X Ser &   244.82347 &     -2.49162 &             5.58 &   J161917.6-022930 &         0.58 &                  0.84 &                0.08 (51) &                 3.6 (51) &                     --  &  $\rm (2.1\pm 0.3)\times 10^{32}$ &  $\rm (2.5\pm 0.4)\times 10^{32}$ \\
    46 &   21/02/1901 (C) &       GK Per &    52.80107 &     43.90426 &             2.73 &   J033112.3+435415 &         1.66 &                  2.04 &  $\rm 0.34\pm0.04\ (44)$ &   $\rm 0.44\pm0.01\ (G)$ &                1.15 (c) &  $\rm (5.6\pm 0.2)\times 10^{31}$ &  $\rm (8.9\pm 0.3)\times 10^{31}$ \\
    47 &   08/03/1898 (C) &    V1059 Sgr &   285.46137 &    -13.16212 &             5.79 &   J190150.7-130944 &         3.20 &                  1.34 &                0.16 (47) &                 0.6 (47) &                     --  &  $\rm (8.6\pm 1.4)\times 10^{30}$ &  $\rm (1.1\pm 0.2)\times 10^{31}$ \\
    48 &   31/12/1891 (C) &        T Aur &    82.99712 &     30.44636 &             8.98 &   J053159.3+302647 &         2.04 &                  5.18 &  $\rm 0.42\pm0.08\ (44)$ &  $\rm 1.08\pm0.37\ (39)$ &                0.68 (g) &  $\rm (5.1\pm 1.8)\times 10^{30}$ &  $\rm (8.8\pm 3.1)\times 10^{30}$ \\
    49 &   24/11/1876 (C) &        Q Cyg &   325.43234 &     42.83978 &             7.18 &   J214143.8+425023 &         3.45 &                  2.69 &  $\rm 0.26\pm0.06\ (44)$ &   $\rm 1.37\pm0.05\ (G)$ &                     --  &  $\rm (1.2\pm 0.3)\times 10^{31}$ &  $\rm (1.8\pm 0.5)\times 10^{31}$ \\
    50 &   12/05/1866 (C) &        T CrB &   239.87814 &     25.92129 &            10.53 &   J155930.8+255517 &        10.74 &                  0.48 &    $\rm 0.1\pm0.1\ (52)$ &    $\rm 0.9\pm0.2\ (52)$ &                     --  &  $\rm (1.7\pm 0.6)\times 10^{30}$ &  $\rm (2.0\pm 0.8)\times 10^{30}$ \\
    51 &   27/04/1848 (C) &     V841 Oph &   254.87830 &    -12.89175 &             6.27 &   J165930.8-125330 &         5.65 &                  1.50 &  $\rm 0.44\pm0.06\ (44)$ &   $\rm 0.82\pm0.02\ (G)$ &     $\rm 1.3\ (g)$ &  $\rm (9.4\pm 1.8)\times 10^{30}$ &  $\rm (1.6\pm 0.3)\times 10^{31}$ \\
    52 &   01/01/1783 (C) &       WY Sge &   293.18008 &     17.74844 &             9.36 &   J193243.2+174454 &         5.75 &                 10.61 &    $\rm 1.6\pm0.3\ (39)$ &    $\rm 4.2\pm0.4\ (39)$ &                     --  &  $\rm (1.1\pm 0.4)\times 10^{32}$ &  $\rm (3.9\pm 1.3)\times 10^{32}$ \\

\hline
\end{longtable}
{\bf Columns:} (1) - Source number; (2) - Time of the CNe detection; (3) - The name of the CN; (4) - Right ascension; (5) - Declination; (6) - X-ray position error, radius 98\%; (7) - The name of the X-ray source from the SRG/eROSITA catalogue; (8) - The angular distance between the optical CNe and the X-ray source; (9) - Galactic absorption in the direction of the source from HI4PI; (10) - Color excess; (11) - Distance to the CN; (12) – The WD mass; (13) - Observed X-ray luminosity in the 0.3--2.3 keV energy band; (14) - X-ray luminosity in the 0.3--2.3 keV energy band, corrected for absorption. \
References:(1) \citet{2021ATel14704....1M}, (2) \citet{2021ATel14471....1M}, (3) \citet{2021ATel14710....1A}, 
(4) \citet{2019ATel13241....1W}, (5) \citet{2020AN....341..781J}, (6) \citet{2019ATel13047....1S}, 
(7) \citet{2020MNRAS.499.4814P}, (8) \citet{2018ATel11928....1W}, (9) \citet{2018ATel11940....1M}, 
(10) \citet{2018ATel11588....1W}, (11) \citet{2020A&A...639L..10M}, (12) \citet{2017ATel10527....1K}, 
(13) \citet{2017ATel10572....1M}, (14) \citet{2018MNRAS.476.4162O}, (15) \citet{2016ATel.9669....1S}, 
(16) \citet{2017NatAs...1..697L}, (17) \citet{2013ATel.5279....1D}, (18) \citet{2019ApJ...878...28S}, 
(19) \citet{2012CBET.3140....2A}, (20) \citet{2011IAUC.9247....3K}, (21) \citet{2008IAUC.8934....1N}, 
(22) \citet{2019ApJS..242...18H}, (23) \citet{2007IAUC.8832....3S}, (24) \citet{2007IAUC.8884....2R}, 
(25) \citet{2008NewA...13..557P}, (26) \citet{2007IAUC.8821....4H}, (27) \citet{2007IAUC.8848....1M}, 
(28) \citet{2009AN....330...77P}, (29) \citet{2006CBET..399....1H}, (30) \citet{2011ApJS..197...31S}, 
(31) \citet{2011ApJS..197...31S}, (32) \citet{2002IAUC.7975....1H}, (33) \citet{2007AAS...210.0402V}, 
(34) \citet{1998IAUC.6941....1N}, (35) \citet{2000ApJ...541..791L}, (36) \citet{1992IAUC.5454....1C}, 
(37) \citet{1988IAUC.4570....2H}, (38) \citet{1976IAUC.2997....1M}, (39) \citet{2016MNRAS.461.1177O}, 
(40) \citet{1995MNRAS.276..353S}, (41) \citet{1975IAUC.2826....1H}, (42) \citet{1971IAUC.2340....1K}, 
(43) \citet{1967IAUC.2022....1C}, (44) \citet{2013A&A...560A..49S}, (45) \citet{1994MNRAS.266..761W}, 
(46) \citet{2002PASP..114..748H}, (47) \citet{1997ApJ...487..226S}, (48) \citet{2016MNRAS.462.1371T}, 
(49) \citet{2004AJ....127..460H}, (50) \citet{2003AJ....126.3017T}, (51) \citet{2004BaltA..13...93S}, 
(52) \citet{2010ApJS..187..275S}. (G) -- Catalogue of Gaia v.2; (С) -- Central Bureau for Astronomical Telegrams. 
WD masses: (a): \citet{2020MNRAS.499.4814P}; (b): \citet{2021gacv.workE..29C}; (c): \citet{2019ApJS..242...18H}; 
(d): \citet{2010NewA...15..657P}; (e): \citet{2010ApJ...709..680H}; (f): \citet{2002ASPC..261..629H}; 
(g): \citet{2007AJ....134.1923P}; (h): \citet{2007ApJ...667..442S}.

\end{landscape}

\end{document}